\documentclass{article}

\usepackage{PRIMEarxiv}

\usepackage[utf8]{inputenc} 
\usepackage[T1]{fontenc}    
\usepackage{hyperref}       
\usepackage{url}            
\usepackage{booktabs}       
\usepackage{amsfonts}       
\usepackage{nicefrac}       
\usepackage{microtype}      
\usepackage{lipsum}
\usepackage{fancyhdr}       
\usepackage{graphicx}       
\graphicspath{{media/}}     
\usepackage{graphicx}%
\usepackage{multirow}%
\usepackage{amsmath,amssymb,amsfonts}%
\usepackage{amsthm}%
\usepackage{mathrsfs}%
\usepackage[title]{appendix}%
\usepackage{xcolor}%
\usepackage{textcomp}%
\usepackage{manyfoot}%
\usepackage{booktabs}%
\usepackage{algorithm}%
\usepackage{algorithmicx}%
\usepackage{algpseudocode}%
\usepackage{listings}%
\usepackage{amsmath} 
\usepackage[table]{xcolor}
\usepackage{multirow}
\usepackage{booktabs}
\usepackage{pdflscape}  
\usepackage{lipsum}    
\usepackage{colortbl}
\usepackage{rotating}
\usepackage{adjustbox}
\newcommand{\cofirst}{\textsuperscript{*}}
\newcommand{\corrauth}{\textsuperscript{\dag}}
\title{BAAI Cardiac Agent: An intelligent multimodal agent for automated reasoning and diagnosis of cardiovascular diseases from cardiac magnetic resonance imaging
}

\author{
\begin{minipage}{\textwidth}
\centering
Taiping Qu$^{1}$\cofirst,
Hongkai Zhang$^{2}$\cofirst,
Lantian Zhang$^{1}$,
Can Zhao$^{1}$,
Nan Zhang$^{2}$,
Hui Wang$^{2}$, \\
Zhen Zhou$^{2}$,
Mingye Zou$^{1}$,
Kairui Bo$^{2}$,
Pengfei Zhao$^{1}$,
Xingxing Jin$^{3}$,
Zixian Su$^{1}$,\\
Kun Jiang$^{2}$, 
Huan Liu$^{1}$,
Yu Du$^{4}$,
Maozhou Wang$^{5}$,
Ruifang Yan$^{3}$\corrauth,
Zhongyuan Wang$^{1}$\corrauth, \\
Tiejun Huang$^{1}$\corrauth,
Lei Xu$^{2}$\corrauth,
Henggui Zhang$^{1}$\corrauth \\[6pt]
\footnotesize
$^{1}$Beijing Academy of Artificial Intelligence, No.\,150 Chengfu Road, Haidian District, Beijing 100084, China \\
$^{2}$Department of Radiology, Beijing Anzhen Hospital, Beijing Institute of Heart, Lung \& Vascular Diseases, Capital Medical University, 2 Anzhen Road, Beijing 100029, China \\
$^{3}$Department of MR, the First Affiliated Hospital, Henan Medical University, 88 Jiankang Road, Weihui 453100, China \\
$^{4}$Department of Cardiology, Clinical Center for Coronary Heart Disease, Beijing Institute of Heart, Lung and Blood Vessel Disease, Beijing Anzhen Hospital, Capital Medical University, Beijing 100029, China \\
$^{5}$Department of Cardiac Surgery, Beijing Anzhen Hospital, Institute of Heart, Lung and Vascular Diseases, Capital Medical University, Beijing 100029, China \\[4pt]
\cofirst These authors contributed equally to this work. \\
\corrauth Corresponding authors: \url{yrf718@163.com}, \url{zhongyuan@baai.ac.cn}, \url{tjhuang@baai.ac.cn}, \url{leixu2001@hotmail.com}, \url{henggui.zhang@gmail.com}
\end{minipage}
}

\begin{document}
\maketitle

\begin{abstract}{Cardiac magnetic resonance (CMR) is a cornerstone for diagnosing cardiovascular disease. However, it remains underutilized due to complex, time-consuming interpretation across multi-sequences, phases, quantitative measures that heavily reliant on specialized expertise. Here, we present BAAI Cardiac Agent, a multimodal intelligent system designed for end-to-end CMR interpretation. The agent integrates specialized cardiac expert models to perform automated segmentation of cardiac structures, functional quantification, tissue characterization and disease diagnosis, and generates structured clinical reports within a unified workflow. Evaluated on CMR datasets from two hospitals (2413 patients) spanning 7-types of major cardiovascular diseases, the agent achieved an area under the receiver-operating-characteristic curve exceeding 0.93 internally and 0.81 externally. In the task of estimating left ventricular function indices, the results generated by this system for core parameters such as ejection fraction, stroke volume, and left ventricular mass are highly consistent with clinical reports, with Pearson correlation coefficients all exceeding 0.90. The agent outperformed state-of-the-art models in segmentation and diagnostic tasks, and generated clinical reports showing high concordance with expert radiologists (six readers across three experience levels). By dynamically orchestrating expert models for coordinated multimodal analysis, this agent framework enables accurate, efficient CMR interpretation and highlights its potentials for complex clinical imaging workflows. Code is available at \href{https://github.com/plantain-herb/Cardiac-Agent}{https://github.com/plantain-herb/Cardiac-Agent}.}
\end{abstract}




\maketitle

\section{Introduction}\label{sec1}

Cardiovascular disease (CVD) remains the leading cause of mortality and morbidity worldwide, imposing an escalating burden on healthcare systems despite advances in prevention and treatment \cite{McNamara2019,Alhabeeb2020, 2014ESC,wang2024screening}. Accurate phenotyping of cardiac structure, function, and tissue pathology is central to effective diagnosis, risk stratification, and therapeutic decision-making across the spectrum of CVD. Among available imaging modalities, cardiac magnetic resonance imaging (CMR) is one of the most comprehensive imaging modalities for non-invasively evaluating cardiovascular diseases, providing  comprehensive, multi-parametric characterization of the heart, encompassing anatomical structure, ventricular function, myocardial perfusion, and tissue composition \cite{2013Advances,2010Quantification}. As a result, CMR is a cornerstone for the assessment of cardiac morphology and function in many clinical contexts \cite{20212D,2020Value}. 

However, the very richness that underpins the clinical value of CMR also presents a fundamental challenge. Contemporary CMR examinations consist of multiple three-dimensional imaging sequences acquired across the cardiac cycle (i.e., 4D sequence: 3D+time), yielding data that are intrinsically spatiotemporal and heterogeneous. Interpreting these data requires the integration of tissue-specific structure and imaging signature, cine motion and quantitative functional indices, a process that is cognitively demanding, time-consuming, and heavily reliant on expert experience \cite{2007Guidelines,hartung2011magnetic}. In routine clinical practice, comprehensive CMR interpretation often requires prolonged manual interaction, careful cross-referencing across sequences, and iterative reasoning steps, which together limit throughput and introduce variability \cite{2007Guidelines,schulz2013standardized}. Moreover, the extensive training required to achieve proficiency in CMR interpretation restricts scalability and contributes to disparities in access to high-quality cardiac imaging expertise \cite{2007Guidelines,hartung2011magnetic}.  

Artificial intelligence (AI) has emerged as a promising avenue to alleviate these constraints \cite{shiri2025ai,wang2024screening}. Deep learning approaches have demonstrated strong performance in individual CMR and other imaging modal tasks, such as cardiac segmentation \cite{QIU2023102694}, functional parameter quantification \cite{ruijsink2020fully}, and disease classification \cite{shiri2025ai}. These task-specific models can rapidly extract objective imaging features and have achieved accuracy comparable to expert readers in controlled settings \cite{Wang2024, Merlo2023, doi:10.1161/CIRCIMAGING.123.016115}. Nevertheless, such approaches remain fundamentally limited in clinical practice because they operate in isolation. CMR interpretation is not a collection of independent sub-tasks but a coordinated reasoning process in which structural, functional, and tissue-level feature information must be jointly considered to arrive at a clinically meaningful conclusion. Fragmented AI solutions fail to capture this integrative logic and therefore struggle to deliver reliable, end-to-end clinical value.

Recent advances in large multimodal models (LMMs) have opened new possibilities for medical image understanding by enabling the joint processing of visual and textual information \cite{li2025towards,Nath_2025_CVPR,pmlr-v225-moor23a,li2023llavamedtraininglargelanguageandvision,HU2024103279,LIANG2024103805}. In radiology, LMMs have shown encouraging capabilities in tasks such as visual question answering (VQA), medical report generation (MRG) \cite{li2025towards,Nath_2025_CVPR}, and multi-modal reasoning \cite{pmlr-v225-moor23a,li2023llavamedtraininglargelanguageandvision,HU2024103279,LIANG2024103805,Nath_2025_CVPR}, and image segmentation \cite{10968301}. However, existing LMMs are primarily developed and evaluated on two-dimensional images or single-slice volumetric data. Extending these models to CMR is non-trivial, as CMR data encode 4D dynamical information, i.e., complex three-dimensional anatomy coupled with dynamic motion across time. Effective interpretation therefore requires not only visual recognition but also spatiotemporal reasoning and domain-specific quantitative analysis, capabilities that remain beyond the scope of current general-purpose LMMs.

A promising strategy to bridge this gap is the development of intelligent agent systems \cite{li2024mmedagentlearningusemedical,NEURIPS2024_90d1fc07,VILLA2025100274,Qiu2024,Mehandru2024} that combine the complementary strengths of specialized expert models and LMMs. In this paradigm, expert models provide precise, high-fidelity quantitative measurements from complex imaging data, while LMMs orchestrate task execution, integrate heterogeneous outputs, and generate clinically coherent interpretations aligned with radiological reasoning. Rather than replacing expert knowledge, such agents aim to emulate the workflow of experienced clinicians by coordinating perception, measurement, and inference in a unified framework. Despite initial successes in limited radiological settings, existing agent-based approaches have largely focused on static images and have not addressed the distinctive challenges posed by 4D spatiotemporal CMR data.

\begin{figure*}[!t]
	\centering
		\includegraphics[width=1 \textwidth] {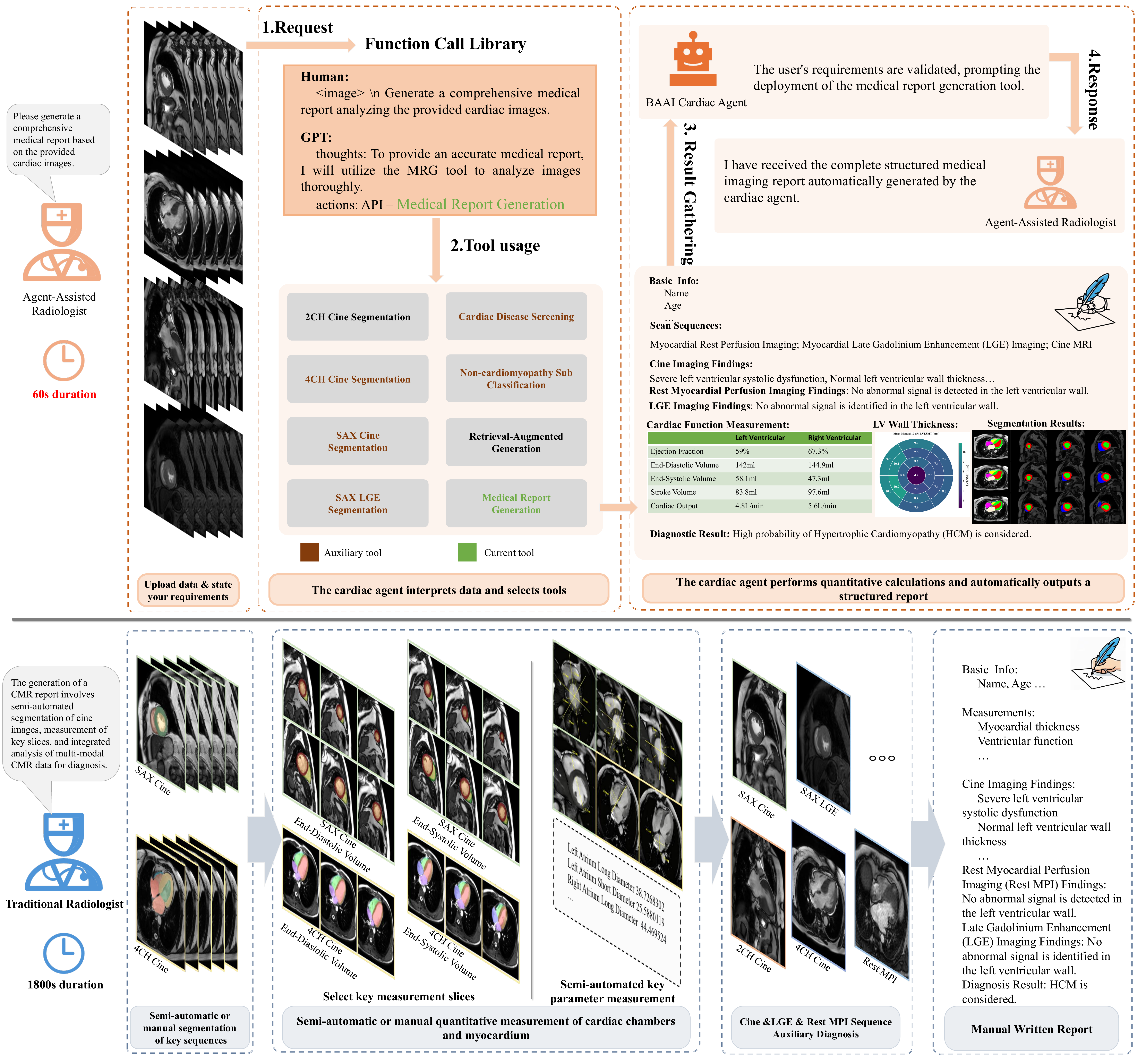} 
	\caption{A schematic overview of the BAAI Cardiac Agent illustrating the process of radiology report generation from multi-sequence CMR scans (upper section), in comparison with the traditional clinical workflow (lower section). 
    Upper section: the agent-assisted workflow, in which the user submits instructions and upload all CMR sequences to the BAAI Cardiac Agent. The agent automatically interprets the request, invokes appropriate analytic modules, and performs segmentation, quantitative assessment, and diagnostic reasoning. A structured, quantitative enriched radiology report is then generated  within approximately 90 seconds.
    Lower section: the workflow of a traditional radiologist. This workflow begins with manual or semi-automatic segmentation of cine images, followed by qualification of key cardiac function parameters. Diagnostic interpretation is then performed by integrating information from late gadolinium enhancement (LGE) and rest myocardium perfusion imaging (Rest MPI), after which the final report is manually compiled. This sequential workflow typically requires approximately 1800 seconds. 
    } \label{fig:1}
\end{figure*}

Here we proposed a  Cardiac  Agent  system (BAAI Cardiac Agent) designed to address key unmet needs in clinical CMR practice (see Fig. \ref{fig:1}). The system has the following major features: 
1) First, it supports the unified variant expert models of segmentation of short-axis (SAX) cine segmentation (SAXCS), two-chamber (2CH) cine segmentation (2CHCS), four-chamber (4CH) cine segmentation (4CHCS), SAX late gadolinium enhancement (LGE) segmentation (SAXLGES) imagings, cardiac disease screening (CDS), non-ischemic cardiomyopathy subclassification (NICMS), interpretation of multiple CMR sequences, as well as retrieval-augmented generation (RAG) and MRG, enabling synchronized reasoning across anatomical, functional, and tissue-specific information. 
2) Second, it integrates quantitative measurements directly into structured, narrative reports, reducing manual effort and improving reproducibility. 
3) Third, it provides interactive capabilities, allowing clinicians to query imaging findings and reasoning steps in a transparent manner. Through this design, the BAAI Cardiac Agent transforms CMR analysis from a labor-intensive, fragmented workflow into a rapid, standardized, and explainable decision-support process.

We validate the BAAI Cardiac Agent on a large-scale, clinically curated CMR dataset with a cohort of 2413 patients from two hospitals, and demonstrate robust performance across segmentation, disease screening, cardiomyopathy subclassification, and quantitative functional assessment. Comprehensive evaluation shows that the agent achieves state-of-the-art (SOTA) accuracy in core imaging tasks while producing measurements and diagnostic conclusions that are highly consistent with expert radiology reports. Importantly, the system substantially reduces interpretation time without compromising clinical fidelity, highlighting its potential to improve efficiency and accessibility in real-world CMR practice.

Collectively, this study establishes a generalizable framework for intelligent, agent-based interpretation of spatiotemporal medical imaging and demonstrates its practical value in one of the most complex domains of radiology. By integrating expert-level quantification with multimodal reasoning, the BAAI Cardiac Agent represents a step toward scalable, trustworthy AI systems that operate in close alignment with clinical expertise and workflow.

\section{Results}\label{sec3}
\subsection{Patient datasets}\label{subsec31}
In this study, CMR scan sequences of 2413 patients were retrospectively collected from two hospitals. Detailed information on data acquisition and institutional sources is provided in the Appendix \ref{secA12}, while summary statistics and demographic characteristics of the datasets are provided in Supplementary Table \ref{demographics}.

\subsection{Evaluation of the Segmentation Expert Model}\label{subsec32}
For segmentation of SAX cine, 2CH cine, 4CH cine, and SAX LGE sequences, 150 samples were randomly selected from the subject cohort for data annotation (Appendix \ref{secA4}). Following stringent quality control, the annotated data were split into training, validation, and test sets in a 7:1:2 ratio. Dedicated segmentation expert models were then developed for each sequence of the SAX cine, 2CH cine, 4CH cine, and SAX LGE types.

To evaluate the performance of the segmentation expert model integrated within the BAAI Cardiac Agent, we conducted fair comparative experiments against several widely used SOTA methods, including nnUNet  \cite{isensee2021nnu}, MedSAM2 \cite{ma2025medsam2}, ResUNet++ \cite{jha2019resunet++}, and DiffUNet \cite{xing2025diff}. Segmentation performances across different CMR sequence types was assessed using three standard metrics: Dice Similarity Coefficient (DSC), Hausdorff Distance (HD), and Average Surface Distance (ASD). Details of the calculation formulas of these metrics are provided in the Appendix \ref{secAcal}.

As summarized in Table \ref{segres}, the proposed segmentation expert model consistently outperforms existing mainstream SOTA methods in cardiac CMR sequence segmentation tasks. It achieves the highest DSC values for all evaluated sequences, while also yielding the lowest HD values in three sequences and the lowest ASD value in two sequences. Across all metrics, it exhibits generally small standard deviations, indicating stable and robust performance. In contrast, the competing methods exhibit only sequence- or metric-specific strengths. For instance, MedSAM2 \cite{ma2025medsam2} performs relatively well in terms of ASD for the 2CH cine sequence and achieves the best HD performance for SAX LGE images, while nnUNet \cite{isensee2021nnu} shows favorable ASD performance for SAX LGE images, but its overall performance remains inferiors to that of the proposed model. DiffUNet \cite{xing2025diff} shows substantial higher HD and ASD values, likely due to noise-based initialization mechanism, which introduces additional fluctuations to boundaries, adversely affecting the performance. 

Quantitative comparisons are presented in Supplementary Fig. \ref{s1}. Compared with other competing methods, the proposed method demonstrates the best overlap with manual annotations across all CMR sequences. Distinct gaps exist between individual cardiac structures, with no obvious structural missing or false positives. Detailed descriptions of the specific evaluation metrics for different cardiac structures in the four CMR segmentation datasets are provided in Supplementary Tables \ref{segressax}, \ref{segres2ch}, \ref{segres4ch} and \ref{segreslge}. Furthermore, clinical practice places particular emphasis on the localization and assessment of LGE involvement extent, and we validate and elaborate on the relevant results of quantitative LGE analysis in Supplementary Fig. \ref{LGEBullseye}.

\begin{table}[htbp]
  \centering
  \small 
  \begin{tabular}{clcccc}
    \toprule
    Metric & Method & SAX cine & 2CH cine & 4CH cine & SAX LGE \\
    \midrule
    \rowcolor{gray!15}
    \multicolumn{1}{c}{\cellcolor{white}}
    & ResUNet++ & 84.69±0.05 & 86.97±0.04 & 84.41±0.06 & 58.08±0.16 \\
    & nnUNet & 87.42±0.06 & 88.14±0.05 & 81.61±0.14 & 73.89±0.18 \\
    \rowcolor{gray!15}
    \multicolumn{1}{c}{\cellcolor{white}DSC}
    & DiffUNet & 77.19±0.13 & 83.05±0.08 & 77.38±0.11 & 47.28±0.20 \\
    & MedSAM2 & 80.86±0.20 & 77.89±0.20 & 82.07±0.08 & 69.70±0.21 \\
    \rowcolor{gray!15}
    \multicolumn{1}{c}{\cellcolor{white}}
    & Ours & \textbf{90.21±0.02} & \textbf{88.75±0.03} & \textbf{86.92±0.03} & \textbf{75.07±0.07} \\
    \midrule
    & ResUNet++ & 16.58±8.13 & 12.28±6.62 & 13.73±10.46 & 16.11±5.57 \\
    \rowcolor{gray!15}
    \multicolumn{1}{c}{\cellcolor{white}}
    & nnUNet & 13.68±9.76 & 29.88±49.74 & 12.17±3.88 & 12.46±5.52 \\
    HD & DiffUNet & 92.30±24.83 & 117.35±36.61 & 72.80±52.81 & 185.59±63.24 \\
    \rowcolor{gray!15}
    \multicolumn{1}{c}{\cellcolor{white}}
    & MedSAM2 & 19.39±5.43 & 9.22±3.81 & 7.94±1.89 & \textbf{12.18±4.49} \\
    & Ours & \textbf{8.24±2.52} & \textbf{7.87±2.18} & \textbf{7.52±1.87} & 14.05±4.89 \\
    \midrule
    \rowcolor{gray!15}
    \multicolumn{1}{c}{\cellcolor{white}}
    & ResUNet++ & 0.94±0.58 & 0.64±0.42 & 0.94±1.27 & 1.29±1.47 \\
    & nnUNet & 0.75±0.78 & 2.72±7.96 & 0.67±0.62 & \textbf{0.63±0.26} \\
    \rowcolor{gray!15}
    \multicolumn{1}{c}{\cellcolor{white}ASD}
    & DiffUNet & 2.65±1.85 & 5.46±5.80 & 8.26±8.12 & 56.65±60.78 \\
    & MedSAM2 & 2.45±0.79 & \textbf{0.25±0.17} & \textbf{0.26±0.11} & 1.15±1.48 \\
    \rowcolor{gray!15}
    \multicolumn{1}{c}{\cellcolor{white}}
    & Ours & \textbf{0.64±0.29} & 0.27±0.17 & \textbf{0.26±0.13} & 1.14±0.78 \\
    \bottomrule
  \end{tabular}
  \caption{Performance comparison between the BAAI Cardiac Agent segmentation expert model and SOTA segmentation networks. Computed DSC (\%), HD, and ASD values are shown as mean ± standard deviation.}
  \label{segres}
\end{table}

\subsection{Evaluation of the CVDs Diagnostic Expert Model
}\label{subsec33}
To ensure reliable model training and unbiased performance evaluation for the CDS and NICMS tasks, the internal dataset was partitioned into training, internal validation set, and internal test sets in a 7:1:2 ratio using a stratified split strategy that strictly follows the number distribution of CVDs categories. The training set was used for iterative model training, the validation set for real-time generalization evaluation and optimal stopping-point determination, and the internal test set for post-training performance assessment. In addition, an independent external validation set was employed to further evaluate the model’s generalization capability.

\subsubsection{Evaluation of CDS}\label{subsec331}
The CDS model utilizes three combined views of SAX cine, 2CH cine, and 4CH cine sequences, and was evaluated on the internal test set (n=365). For normal heart (NH) screening, the model achieved an area under the curve (AUC) of 0.980 (95\% confidence interval [CI]: 0.956–0.995) and an F$_1$-score of 0.856 (95\% CI: 0.782–0.914). In  anomaly detection tasks, ischemic heart disease (IHD) screening yielded an AUC of 0.938 (95\% CI: 0.911–0.961), an F$_1$-score of 0.860 (95\% CI: 0.819–0.897), with a sensitivity of 0.820 (95\% CI: 0.760–0.875) and a specificity of 0.931 (95\% CI: 0.897–0.962). For non-ischemic cardiomyopathy (NICM) screening, the model attained an AUC of 0.960 (95\% CI: 0.943–0.975), an F$_1$-score of 0.857 (95\% CI: 0.810–0.896), a sensitivity of 0.879 (95\% CI: 0.821–0.932), and a specificity of 0.893 (95\% CI: 0.852–0.931). 

\begin{figure*}[!t]
	\centering
		\includegraphics[width=1 \textwidth] {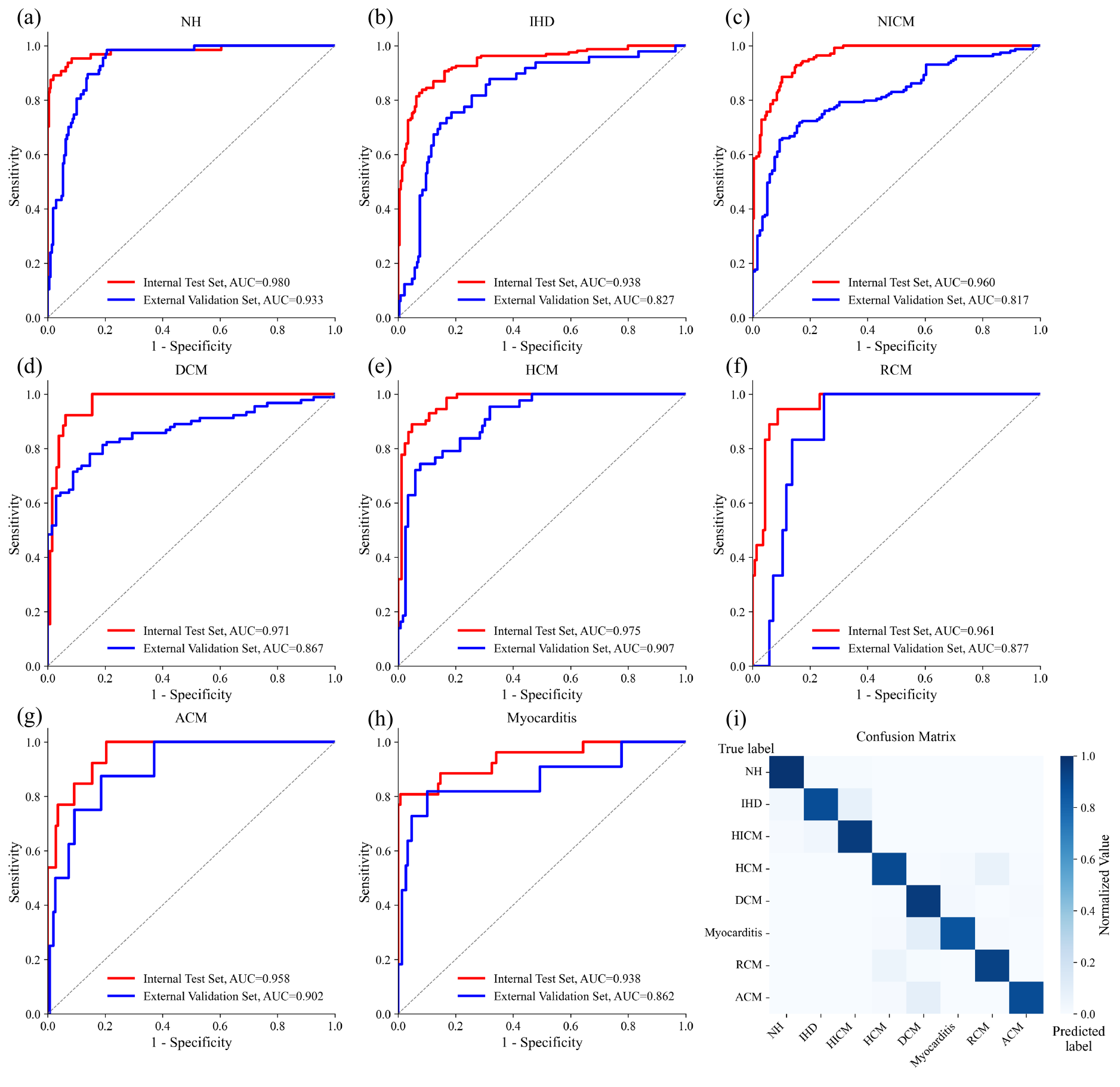} 
	\caption{Performance of the expert CVDs diagnostic model in internal and external testing. a–c: Receiver operating characteristic curves for CDS; d–h: Receiver operating characteristic curves for NICMS. The red curves represent the internal test set, and the blue curves represent the external validation set. i: Confusion matrix of the AI diagnostic model predictions versus ground truth labels based on the full cohort (n = 2,232), with a color gradient visually representing the proportional distribution of predictions across disease categories.
    } \label{figcls}
\end{figure*}

External validation on an independent cohort (n=275) demonstrated stable performance, with AUC values exceeding 0.810 across all three categories, supporting the model's generalization ability. Detailed results are presented in Fig. \ref{figcls} (a–c) and Table \ref{cc}. Notably, the study cohort comprises multiple types of CVDs (Table \ref{demographics}), indicating the robustness of the screening model across diverse disease types.

\begin{table*}[htbp]
  \centering
  \renewcommand{\arraystretch}{1}
  \setlength{\tabcolsep}{4pt}
  \rowcolors{4}{gray!15}{white} 
  \begin{tabular}{lcccccccccc}
    \hline
    &\multicolumn{5}{c}{Internal Test Set} & \multicolumn{5}{c}{External Validation Set} \\
    \cmidrule(lr){2-6} \cmidrule(lr){7-11}
    & N & Ours & VST & ViT & ResNet & N & Ours & VST & ViT & ResNet \\
    NH & 64 & 0.856 & 0.839 & 0.855 & \textbf{0.864} & 67 & \textbf{0.753} & 0.712 & 0.720 & 0.717 \\
    IHD & 161 & \textbf{0.860} & 0.812 & 0.799 & 0.835 & 49 & 0.582 & 0.601 & 0.544 & \textbf{0.631} \\
    NICM & 140 & \textbf{0.857} & 0.830 & 0.817 & 0.832 & 159 & \textbf{0.762} & 0.716 & 0.703 & 0.713 \\
    F-W F$_1$ &  & \textbf{0.858} & 0.824 & 0.817 & 0.839 &  & \textbf{0.728} & 0.694 & 0.679 & 0.699 \\
    Accuracy &  & \textbf{0.858} & 0.825 & 0.816 & 0.833 &  & \textbf{0.724} & 0.687 & 0.673 & 0.695 \\
    \bottomrule
  \end{tabular}
  \caption{Performance comparison of cardiac disease screening (CDS) in normal heart (NH), ischemic heart disease (IHD) and non-ischemic cardiomyopathy (NICM) cohorts between the BAAI Cardiac Agent CDS expert model and three SOTA models. (N: Number of subjects, F-W F$_1$: Frequency-weighted F$_1$)}
  \label{cc}
\end{table*}

\subsubsection{Evaluation of NICMS}\label{subsec332}
The diagnostic model for NICMS adopts the combined input of dual-view cine sequences (SAX, 4CH) and SAX LGE, thereby leveraging the complementary structual, functional, and tissue-characterization information provided by CMR. On the internal test set (n=155), the model achieved a class-weighted mean AUC of 0.9650 (95\% CI: 0.944-0.998) and an F$_1$-score of 0.834 (95\% CI: 0.779–0.890). 

Across all disease categories, AUC values exceeded 0.930, and F$_1$-scores were greater than 0.800 for all subclasses except restrictive cardiomyopathy (RCM) and arrhythmogenic cardiomyopathy (ACM). Notably, the model demonstrated excellent performance for the most prevalent NICM subtypes, achieving an AUC of 0.975 (95\% CI: 0.948–0.992) and F$_1$-score of 0.899 (95\% CI: 0.844–0.947) for hypertrophic cardiomyopathy (HCM), and an AUC of 0.971 (95\% CI: 0.945–0.992) with an F$_1$-score of 0.800 (95\% CI: 0.667–0.895) for dilated cardiomyopathy (DCM). In addition, the model also achieved a high AUC of 0.938 (95\% CI: 0.870–0.991) and an F$_1$-score of 0.857 (95\% CI: 0.722–0.955) for myocarditis. 

External validation on an independent cohort (n=159) confirmed the stability and generalizability of the model, with AUC values exceeding 0.860 for all five NICM categories. Detailed results are presented in Fig. \ref{figcls} (d–h) and Table \ref{ncc}. 

\begin{table*}[htbp]
  \centering
  \renewcommand{\arraystretch}{1.0}
  \setlength{\tabcolsep}{4pt}
  \rowcolors{4}{gray!15}{white} 
  \begin{tabular}{lcccccccccc}
    \hline
    &\multicolumn{5}{c}{Internal Test Set} & \multicolumn{5}{c}{External Validation Set} \\
    \cmidrule(lr){2-6} \cmidrule(lr){7-11}
    & N & Ours & VST & ViT & ResNet & N & Ours & VST & ViT & ResNet \\
    HCM & 72 & \textbf{0.899} & 0.865 & 0.848 & 0.888 & 43 & \textbf{0.769} & 0.729 & 0.719 & 0.748 \\
    DCM & 26 & 0.800 & 0.798 & \textbf{0.813} & 0.800 & 91 & \textbf{0.812} & 0.711 & 0.721 & 0.714 \\
    RCM & 18 & 0.732 & 0.762 & \textbf{0.815} & 0.762 & 6 & 0.444 & 0.483 & \textbf{0.567} & \textbf{0.567} \\
    ACM & 13 & 0.636 & 0.636 & \textbf{0.727} & \textbf{0.727} & 8 & 0.421 & \textbf{0.657} & 0.435 & 0.421 \\
    Myocarditis & 26 & \textbf{0.857} & 0.779 & 0.779 & 0.797 & 11 & 0.621 & \textbf{0.671} & 0.597 & 0.653 \\
    F-W F$_1$ &  & \textbf{0.834} & 0.807 & 0.818 & 0.830 &  & \textbf{0.754} & 0.702 & 0.691 & 0.698 \\
    Accuracy &  & \textbf{0.832} & 0.807 & 0.813 & 0.819 &  & \textbf{0.730} & 0.685 & 0.667 & 0.673 \\
    \bottomrule
  \end{tabular}
  \caption{Performance comparison between the BAAI Cardiac Agent expert diagnostic model and three SOTA models in the subclassification of non-ischemic cardiomyopathy, including hypertrophic cardiomyopathy (HCM), dilated cardiomyopathy (DCM), restrictive cardiomyopathy (RCM), arrhythmogenic cardiomyopathy (ACM) and myocarditis.}
  \label{ncc}
\end{table*} 

\subsubsection{Comparison of CVDs Diagnosis with Other Models}\label{subsec333}
To assess the relative performance of the proposed diagnostic expert model within the BAAI Cardiac Agent, we conducted fair comparative experiments against representative mainstream backbone networks in the fields of computer vision and medical image analysis, including Video-based Swin Transformer (VST) \cite{wang2024screening}, Vision Transformer (ViT) \cite{dosovitskiy2020image}, and ResNet \cite{qiu2017learning}. Detailed quantitative comparisons among the models are reported in Table \ref{cc} and Table \ref{ncc}. For the CDS task, the proposed model achieved the highest class-weighted F$_1$-score, with superior performance in both IHD and NICM categories. In the NICMS task, the proposed model likewise attained the optimal weighted F$_1$-score, and outperformed competing methods in key subclasses, including HCM and myocarditis. 

These results demonstrate the robust advantages of the proposed model for CDS and its accuracy in identifying subdivided diseases. 

In addition, comprehensive comparison of model performances between the proposed method and three types of SOTA methods in terms of AUC, sensitivity, and specificity metrics for the tasks of CDS and NICMS, and the relevant results were conducted, and the results are shown in Supplementary Tables \ref{AUC}, \ref{sen} and \ref{spc}. These results further demonstrated that the BAAI Cardiac Agent outperformed SOTA methods.

\subsection{Evaluation of Automatic Cardiac Structural and Functional Measurement
}\label{subsec34}
\begin{figure*}[!t]
	\centering
		\includegraphics[width=1 \textwidth] {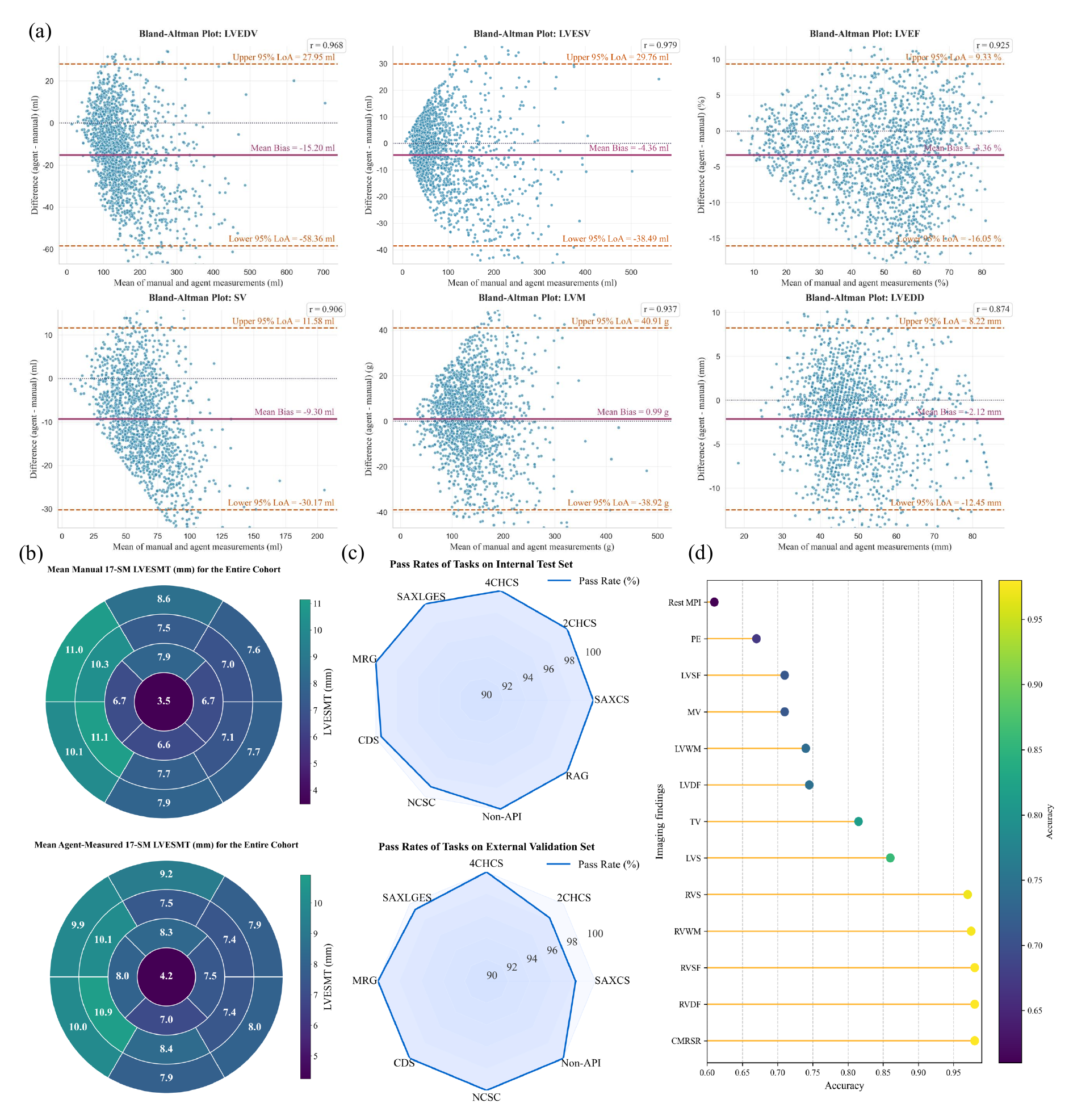} 
	\caption{Evaluation results of BAAI Cardiac Agent report consistency and expert model invocation success rate. (a) Bland-Altman plots comparing BAAI Cardiac Agent measurements (left ventricular end-diastolic volume, LVEDV; left ventricular end-systolic volume, LVESV; left ventricular ejection fraction, LVEF; stroke volume, SV; left ventricular mass, LVM; left ventricular end-diastolic diameter, LVEDD) with manual reports in the internal cohort; (b) Mean distribution of left ventricular end-diastolic wall thickness (LVEDWT) measured by manual reports using the 17-SM and by the BAAI Cardiac Agent in the internal cohort. The bullseye plot correlates the basal, mid, and apical segments of the left ventricle with the outer, middle, and inner layers in sequence, with the apex at the center; (c) Success rate of expert model invocations by the BAAI Cardiac Agent on the internal and external test sets; (d) Accuracy performance of the BAAI Cardiac Agent in cardiac-specific VQA tasks.
    } \label{figmrg}
\end{figure*}

We performed multi-dimensional quantitative analysis of cardiac structural and functional parameters based on the segmentation results of CMR cine sequence images, so as to systematically assess the measurement accuracy and clinical applicability of the BAAI Cardiac Agent. To evaluate the consistency and reliability between the measurement results of the agent and those obtained by manual measurement from clinicians, we also calculated six primary cardiac function parameters based on the segmentation results of SAX cine sequence images, and analyzed the data by using Bland-Altman plots (Fig. \ref{figmrg}(a)). 

The results demonstrated strong concordance between the automated and manual measurements. Pearson correlation coefficients ($r$) of LV end-diastolic volume (LVEDV), and LV end-systolic volume (LVESV) both exceeds 0.960, while the $r$ of LV ejection fraction (LVEF), stroke volume (SV), and LV mass (LVM) were also above 0.900. For LVED diameter (LVEDD), a distance-based measurement parameter, the $r$ is 0.874. These findings indicate excellent correlations in the measurement results of all core cardiac function parameters. The consistency evaluation results of other cardiac function parameters, including cardiac output (CO) and the maximum transverse diameters of the left and right atria (LAT4CHD, RAT4CHD) measured in the 4CH view are presented in Supplementary Fig. \ref{s2} (a). The above results demonstrate that the measurement results of the BAAI Cardiac Agent are highly consistent with those of manual measurement by clinicians, which provides reliable data support for its clinical application and popularization.

Based on the American Heart Association (AHA) 17-Segment Model (17-SM), we analyzed the measurement results of LVED wall thickness (LVEDWT) and visualized the average segmental wall thickness distribution across the internal dataset using polar maps (Fig. \ref{figmrg}(b)). The zonal layout of each polar map strictly adheres to the AHA 17-SM. The visualization results show that mean wall thickness of the anteroseptal wall (ASW) in the basal segment and the septal wall (SW) in the apical segment exceeds 1 mm, while that of all other segments is below 1 mm. 

To further assess measurement fidelity, we compared segment-wise wall thickness between the NH and HCM cohorts of the internal dataset (Supplementary Fig. \ref{s2}(b)). The results demonstrate that the differences between the values measured by the agent and those in clinical reports were minimal across all segments. Notably, in the HCM cohort, the greatest wall thickening was observed in the basal inter-ventricular septum, a region known to experience the highest mechanical load and to be the most common site of myocardial hypertrophy seen in clinical practice. The concordance between these findings and established pathological observations provides additional evidence of the accuracy and clinical validity of the BAAI Cardiac Agent for ventricular structure assessment under pathological conditions.

\begin{table}[htbp]
  \centering
  \begin{tabular}{lcccccc}
    \hline
    \textbf{Model} &
    \multicolumn{2}{c}{\textbf{Junior radiologists}} &
    \multicolumn{2}{c}{\textbf{Mid radiologists}} &
    \multicolumn{2}{c}{\textbf{Senior radiologists}} \\
    \cmidrule(lr){2-3} \cmidrule(lr){4-5} \cmidrule(lr){6-7}
    & \textbf{Score} & \textbf{Conf.} 
    & \textbf{Score} & \textbf{Conf.} 
    & \textbf{Score} & \textbf{Conf.} \\
    \hline
    \rowcolor{gray!15}
    LLaVa-Med     & 26.10±2.83 & 24.50 & 22.07±3.22 & 21.00 & 13.48±3.50 & 13.50 \\
    MedM-VL       & 27.93±6.12 & 23.00 & 23.85±6.81 & 22.00 & 15.65±5.50 & 19.00 \\
    \rowcolor{gray!15}
    MMedAgent     & 24.05±3.20 & 22.50 & 20.02±3.60 & 20.00 & 12.78±3.25 & 17.50 \\
    Qwen-VL-30B   & 58.52±3.73 & 52.50 & 57.18±4.68 & 49.50 & 51.67±5.75 & 46.00 \\
    \rowcolor{gray!15}
    Ours          & \textbf{87.93±2.20} & \textbf{94.00} 
                  & \textbf{87.52±2.84} & \textbf{90.50} 
                  & \textbf{86.53±4.21} & \textbf{87.50} \\
    \hline
  \end{tabular}
  \caption{Performance comparison of different LMMs across radiologist experience levels (junior, middle and senior). Score denotes the comprehensive rating by radiologists of CMR report generation by different models, and Conf. denotes the increase in confidence in writing reports reported by radiologists after reviewing the model output.}
  \label{radiologist_comparison}
\end{table}

\subsection{LMM Performance Evaluation for BAAI Cardiac Agent
}\label{subsec35}
The trained BAAI Cardiac Agent achieves a near-perfect tool selection success rate (close to 100\%), indicating its strong capability to accurately identify and invoke appropriate tools. To rigorously and comprehensive evaluate this ability, we randomly sampled 200 patients from the internal dataset and included all 279 patients from the external dataset for validation. The results show that across diverse CMR sequence tasks, the LMM exhibits robust 3D data comprehension, achieving a tool invocation success rate of 99.82\% on the internal validation dataset and 99.46\% on the external validation dataset. The success rates for individual tools are shown in Fig. \ref{figmrg}(c).

We evaluated the performance of the LMM on VQA tasks for CMR imaging, with results summarized in the lollipop chart (Fig. \ref{figmrg}(d)). Overall, the model achieves an accuracy exceeding 0.965 in assessing the structural normality of categories with low abnormality rates, including right ventricular size (RVS), RV wall motion (RVWM), RV systolic function (RVSF), and RV diastolic function (RVDF). In contrast, for categories with higher abnormality prevalence, such as LVS, LVWM, LVSF, LVDF, mitral valve (MV), tricuspid valve (TV), pericard and rest myocardial perfusion imaging (Rest MPI) abnormalities, the evaluation accuracy ranges from 0.610 to 0.860. 

Using the larger-scale Qwen3-VL-30B \cite{yang2025qwen3} model as a baseline, we conducted a fair comparison across four representative tasks: LVSF assessment, MV analysis, pericardial effusion (PE) evaluation, and Rest MPI. Qwen3-VL-30B achieved accuracy rates of 0.600, 0.380, 0.625, and 0.500 respectively, which are consistently lower than those obtained by the proposed BAAI Cardiac Agent by margins of 0.110, 0.330, 0.045, and 0.110 respectively. These results indicate that the baseline Qwen3-VL-30B \cite{yang2025qwen3} model exhibits limited capability in  detecting CMR abnormalities. Although the screening performance of the proposed BAAI Cardiac Agent remains moderate for the above four tasks, it established a solid foundation for accurate image interpretation and clinical description in large-scale medical multimodal models.

In addition, the model achieves an overall accuracy of 0.980 in the CMR sequence recognition (CMRSR) task. We further conducted qualitative benchmarking against leading open-source large multimodal model (LMM) frameworks including MedM-VL \cite{shi2025medm}, LLaVA-Med \cite{li2023llava}, MMedAgent \cite{li2024mmedagentlearningusemedical} and Qwen3-VL-30B \cite{yang2025qwen3} across a diverse set of tasks. As presented in the Supplementary Fig. \ref{s4}, BAAI Cardiac Agent consistently outperforms these models with more robust and reliable outputs: it not only addresses complex tasks with high accuracy but also achieves zero hallucination, effectively avoiding misleading conclusions such as misidentification of CMR sequences and erroneous calculation of LVEF. Furthermore, its strong generalization capability enables adaptation to diverse CMR scan protocols and heterogeneous patient cohorts, an advantage that existing LMMs constrained by domain-specific data fail to match.

A complete CMR medical report is automatically generated by integrating results from cardiac function measurement, CVDs diagnostic conclusions, and imaging findings. Using the 200 medical reports randomly sampled from the internal dataset described above, we quantitatively evaluated the semantic similarity between the generated reports and the corresponding reference reports using the Bidirectional Encoder Representations from Transformers Score (BERT Score). The evaluation yields a precision, recall, and F$_1$-score of 0.903, 0.894, and 0.898, respectively, indicating a level semantic consistency between the generated and reference reports. A complete example of the CMR report is presented in Supplementary Fig. \ref{s3}. 

To further verify the clinical performance of each LMM and agent, we conduct subjective evaluations on 200 internal patients with complete CMR reports by two radiologists from each of three experience levels: junior (less than 5 years), mid (5–10 years), and senior (more than 10 years). The evaluation uses a 100-point scale, with detailed scoring criteria provided in the Appendix \ref{secA5}. Table \ref{radiologist_comparison} summarizes the corresponding performance results across these groups. CMR report scoring results are presented as mean ± standard deviation. The proposed BAAI Cardiac Agent achieves significantly higher scores and better stability across all three groups. Qwen-VL-30B \cite{yang2025qwen3} shows moderate performance and ranks second, whereas LLaVa-Med \cite{li2023llava}, MedM-VL \cite{shi2025medm}, and MMedAgent \cite{li2024mmedagentlearningusemedical} obtain considerably lower scores, especially under the stricter assessment of senior radiologists. Additionally, the table presents the improvement in report-writing confidence among radiologists at different levels following the use of various models, with a maximum score of 100. Our BAAI Cardiac Agent still achieves the highest rating, followed by Qwen-VL-30B \cite{yang2025qwen3}, while the other comparative models yield relatively smaller gains in radiologist confidence. Overall, radiologists with more experience assign lower ratings, and the BAAI Cardiac Agent consistently outperforms existing mainstream models in clinical evaluation.

\section{Discussion}\label{sec4}
Unlike general medical imaging tasks, cardiac imaging diagnosis constitutes a highly composite clinical workflow that integrates quantitative functional assessment, sequence dependency, temporal phase sensitivity, standardized processing, and clinical interpretability. As the recognized gold standard for evaluating cardiac structure and function, CMR has been integrated into the entire clinical management pathway of CVDs \cite{20212D,2020Value}, including screening, diagnosis, treatment response monitoring, and prognostic evaluation. Particularly in the differential diagnosis of suspected complex CVDs, CMR demonstrates irreplaceable advantages by enabling precise characterization of myocardial tissue features.

Despite the rich diagnostic information offered by multimodal CMR, its potential has not yet been fully translated into clinical efficiency, and conventional CMR analysis remains heavily dependent on physician experience and manual interpretation, leading to persistent challenges. The interpretation of imaging phenotypes is complex, as cardiac motion coupled with multi-sequence data imposes stringent requirement on clinician's knowledge of cardiac anatomy, physiology and pathology \cite{2007Guidelines,hartung2011magnetic}. Manual processing procedures are labor-intensive and time-consuming, involving extensive repetitive quantitative analysis that substantially increase physician workload. Moreover, limited standardization and pronounced subjectivity in analysis hinder reproducibility, restrict inter-institutional collaboration, and compromise the comparability of clinical outcomes.

To directly address the clinical bottlenecks in CMR interpretation, we developed BAAI Cardiac Agent, the first end-to-end agent framework specifically designed for CMR imaging analysis. The framework enables integrated analysis of multi-sequence CMR images, conducting functions such as automatic segmentation of cardiac structures, screening and diagnosis of CVDs, MRG, RAG, and mining of imaging manifestations. Validation based on a large-scale clinical cohort of 2,413 cases demonstrates that BAAI Cardiac Agent delivers robust and comprehensive CMR imaging interpretation capabilities across multiple evaluation dimensions. The system effectively emulates radiologist-level sequence-specific analysis, reliably extracts clinically relevant imaging features, and produces structured reports that are consistent with expert interpretations. By providing an integrated and scalable solution for intelligent cardiovascular imaging analysis, BAAI Cardiac Agent facilitates the translation of CMR from a high-information yet resource-intensive modality into a clinically efficient diagnostic tool, with the potential to streamline imaging workflows and optimize downstream clinical decision-making.


The BAAI Cardiac Agent not only possesses visual question-answering capabilities but also integrates eight expert models, including segmentation expert models for cardiac structure segmentation (SAX, 2CH, 4CH cine, and SAX LGE), the CDS and NICMS models for cardiovascular disease screening and diagnosis, as well as MRG and RAG models. For the tasks of multi-sequence segmentation in CMR and cardiovascular disease screening and diagnosis, this study proposes a generalized CMR segmentation and diagnosis framework. Compared with advanced approaches such as ResUNet++ \cite{jha2019resunet++}, nnUNet \cite{isensee2021nnu}, DiffUNet \cite{xing2025diff}, and MedSAM2 \cite{ma2025medsam2}, the proposed framework achieves the best performance in terms of the DSC across all four sequences and significantly outperforms its counterparts in terms of the HD and ASD metrics. In terms of disease screening and diagnosis, the model achieves an AUC exceeding 0.930 for various cardiovascular diseases on the internal test set and demonstrates strong generalization performance on external test sets. Compared with mainstream models such as ResNet \cite{qiu2017learning}, ViT \cite{dosovitskiy2020image}, and VST \cite{wang2024screening}, our method exhibits clear advantages in both screening and diagnostic effectiveness.

We evaluate the consistency between the key cardiac function parameters automatically measured by the BAAI Cardiac Agent and the clinical imaging reports in the internal dataset. The results show that the correlation coefficients of the LVEDV, LVESV, LVEF, SV, LVM, and LVEDD automatically measured by the Agent with the imaging reports are 0.968, 0.979, 0.925, 0.906, 0.937, and 0.874 respectively, indicating its good clinical applicability. In addition, we further compare the mean predicted myocardial thickness of each segment by the Agent based on the AHA 17-SM with the mean clinical report values in the entire cohort, normal population cohort, and HCM cohort. The results demonstrate that the thickness prediction error remains within 1 mm in each segment of different cohorts, reflecting the stability and anatomical consistency of its measurement results. Overall, the text quality of the generated imaging reports is evaluated by BERT Score, with precision, recall, and F$_1$ score reaching 0.901, 0.896, and 0.898 respectively, verifying from the intelligent text quality perspective that BAAI Cardiac Agent reports are content-accurate, information-complete and clinically reliable. By virtue of a highly integrated, expert model system validated on large-scale clinical cohorts, BAAI Cardiac Agent enables reliable full-chain output ranging from image analysis to report generation, thereby laying a solid technical foundation for the intelligent and standardized interpretation of CMR.

The core engine of the BAAI Cardiac Agent is built upon large-scale medical models with native support for 3D image inputs. In contrast to existing open-source advanced LMM frameworks including MedM-VL \cite{shi2025medm}, LLaVA-Med \cite{li2023llava}, MMedAgent \cite{li2024mmedagentlearningusemedical}, and Qwen3-VL-30B \cite{yang2025qwen3}, the proposed system can deeply understand the spatiotemporal features of 3D cardiac images. Based on 3D inputs, the system can accurately dispatch expert models according to user needs, with task invocation pass rates approaching 100\% on both internal and external test sets. Compared with mainstream LMMs, BAAI Cardiac Agent confers a substantial reliability advantage in the execution of specific tasks, as it can effectively eliminate model hallucinations and achieve accurate understanding of task requirements as well as efficient response.

Moreover, in the task of describing the imaging manifestations of CMR images across different sequences, BAAI Cardiac Agent with 7B parameters consistently outperforms the larger-scale Qwen3-VL-30B model in terms of descriptive  accuracy and clinical relevance. Together, the aforementioned results demonstrate that BAAI Cardiac Agent delivers more accurate, reliable and clinically aligned image interpretation and reporting, while supporting an efficient and automated analytical workflow. This performance highlights its superior comprehensive performance and clinical utility for complex cardiac image analysis.

Naturally, this study has several limitations that warrant to be further addressed in future. First, at the dataset level, although the included CVDs categories encompass most common clinical presentations, the sample size for rare and diagnostically challenging subtypes, particularly ACM and RCM, remains limited due to their low incidence. Consequently, the current data scale does is not yet sufficient to fully support the verification of the model's diagnostic efficacy for such diseases. Future work will expand these cohorts through multi-center cooperation, longitudinal data collection to further improve model robustness and generalizability across full spectrum of CVDs. In addition, the study population are mainly from the East Asian. Given known differences in regional genetic backgrounds, lifestyle factors, and clinical diagnosis and treatment standards, further validation in diverse populations from Europe, the Americas, and Africa will be necessary to establish cross-ethnic diagnostic consistence and support global deployment. 

Second, at the level of diagnostic modality integration, diagnosis of CVDs is a complex process involving multi-dimensional and multi-modal collaboration. Although CMR is recognized as the gold standard for CVD diagnosis and cardiac structural and function assessment, specific diseases phenotypes may be more directly or efficiently characterised using specific imaging modalities, for example, coronary computed tomography angiography (CCTA) for coronary artery disease (CAD) \cite{neglia2023use}, or echocardiography for dynamic assessment of myocardial hypertrophy morphology \cite{ommen20202020}. The current framework focuses primarily on CMR images, and does not yet incorporate additional imaging modalities (such as CT, ultrasound) or non-imaging clinical data (such as electrocardiogram, or laboratory tests). Future development will extend BAAI Cardiac Agent toward a CMR-centered, multimodal diagnostic framework that integrates coronary CTA, echocardiography, and structured clinical information. This evolution reflects not a limitation of the current approach, but a natural progression toward a more comprehensive and clinically aligned cardiovascular intelligence system capable of supporting complex diagnostic reasoning and downstream clinical decision-making.


In summary, this study demonstrates that BAAI Cardiac Agent can emulate key components of expert radiologist practice by performing quantitative cardiac structural and function assessment, detecting cardiac abnormalities, generating structured diagnostic reports, and contextualizing findings with medical knowledge for patient communication. Following rigorous clinical validation, this approach has the potential to substantially improve the efficiency, consistency, and scalability of CMR interpretation, providing a foundation for broader application in CVD and other complex clinical imaging diagnosis.

\section{Method}\label{sec2}
\subsection{Ethics approval and data acquisition}\label{subsec21}
All datasets were collected with approval from the Institutional Review Board (IRB) of Beijing Anzhen Hospital affiliated to Capital Medical University (2025216x), and the First Affiliated Hospital of Xinxiang Medical University (2019164). All data were de-identified prior to model development. Owing to the retrospective nature of this study, the requirement for individual informed consent was waived by the IRBs. Details of CMR acquisition for all patients are provided in the Appendix \ref{secA11}.


\begin{figure*}[!t]
	\centering
		\includegraphics[width=1 \textwidth] {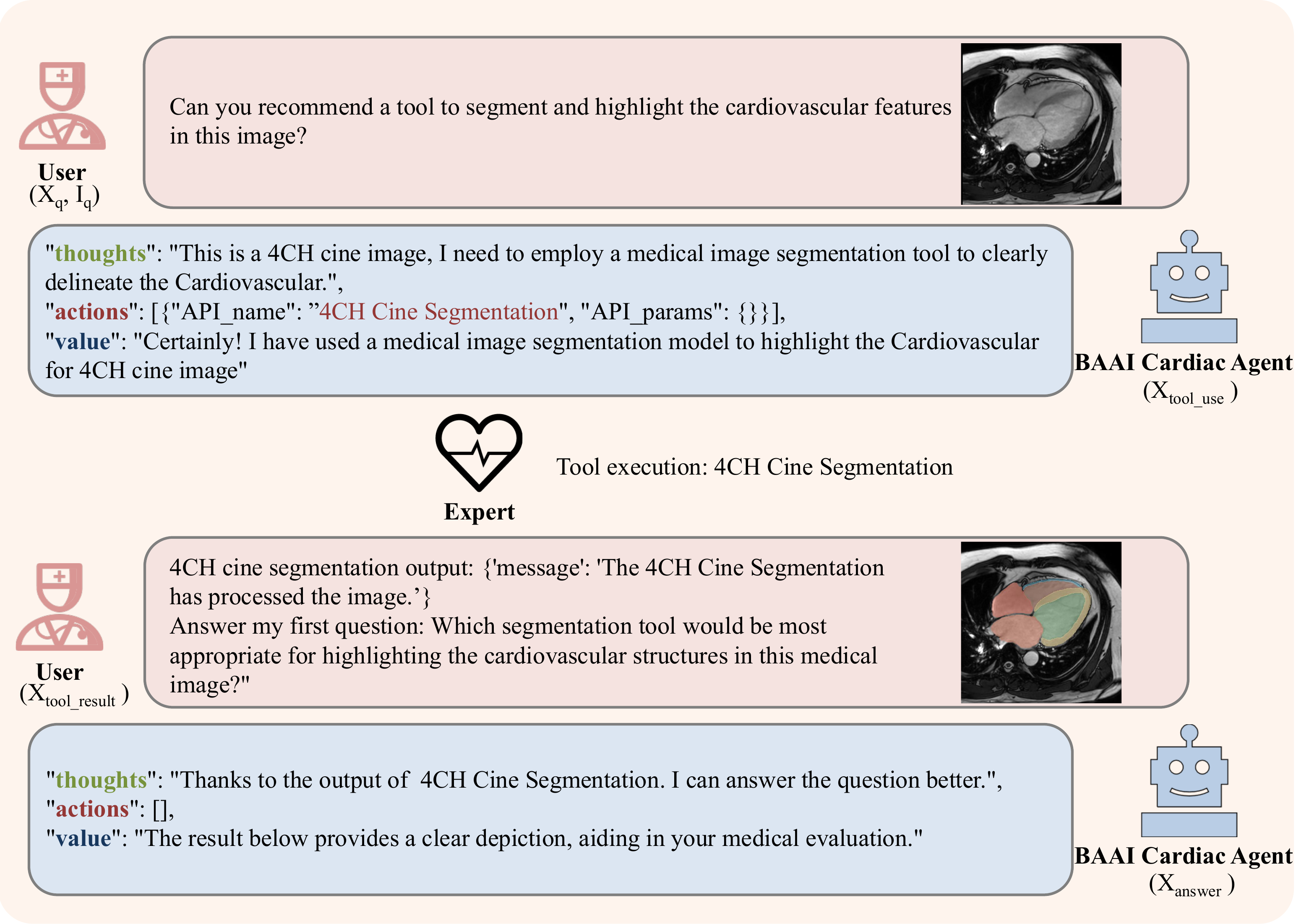} 
	\caption{Schematic diagram of the BAAI Cardiac Agent workflow, in which assigned tasks are completed through adaptive invocation and coordinated use of specialized medical image segmentation tools.} \label{agent_tooluse}
\end{figure*}

\begin{figure*}[!t]
	\centering
		\includegraphics[width=1 \textwidth] {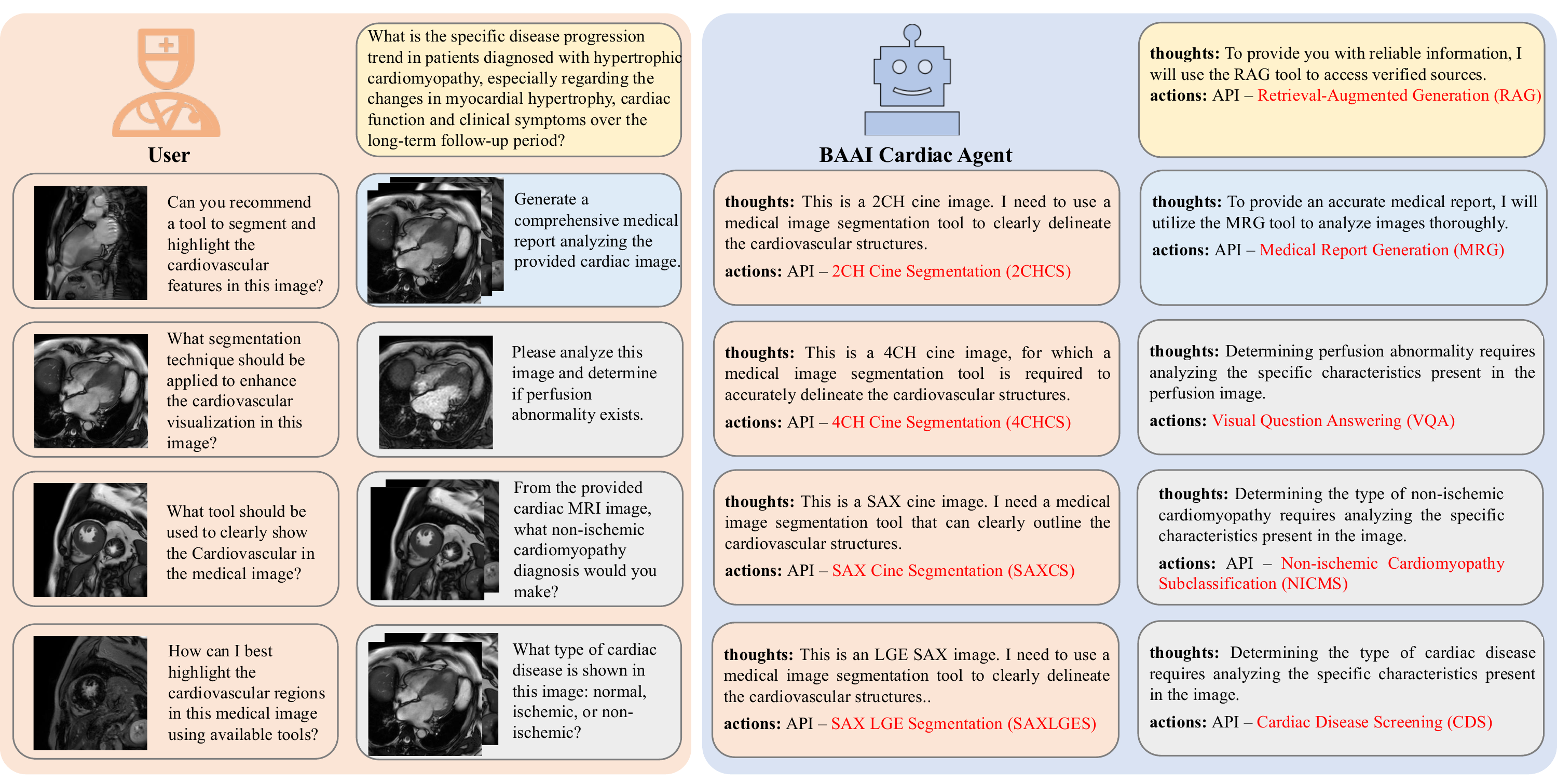} 
	\caption{Task-modality-tool mapping of the BAAI Cardiac Agent: the left panel presents the tasks integrated in the agent and their corresponding medical imaging modalities, while the right panel lists the tools required to complete the respective imaging tasks.} \label{tool_task}
\end{figure*}

\subsection{LMM serves as the engine of the BAAI Cardiac Agent}\label{subsec23}
Following the paradigm of LLaVA-Plus \cite{liu2024llava} and LLaVA-Med \cite{li2023llava}, our objective is to enable LLM to effectively leverage a diverse set of medical multi-modal tools, thereby supporting complex real-world clinical scenarios. In this work, we use the LLaVA \cite{liu2023visual} model as an illustrative example; however, it should be noted that other LMMs are equally applicable as the engine of agents in medical scenarios. 

For an LMM, its workflow typically commences with an image query $\text I_{q}$ input by the user, followed by the reception of a natural language instruction $\text X_{q}$ provided by the user, and ultimately outputs a natural language answer $\text X_{answer}$. Therefore, a unified format can be adopted to represent multimodal instruction-following data, with the specific form presented as follows:
\begin{equation} \label{equ:agent_inp_out}
\texttt{User} : \mathrm{I}_q <\texttt{\textbackslash n}> \mathrm{X}_q <\texttt{STOP}> \quad \texttt{LMM} : \mathrm{X}_{answer} <\texttt{STOP}>
\end{equation}
where  $<\texttt{\textbackslash n}>$ and $<\texttt{STOP}>$ are the newline marker and sequence end marker, respectively. This format is naturally suitable for various multimodal tasks that can be formulated as language-image inputs with language-based outputs. It accommodates a variety of visual tasks, including image recognition, image description, VQA, and MRG, among others.


\subsubsection{BAAI Cardiac agent workflow}\label{subsubsec231}
To ensure the BAAI Cardiac Agent simultaneously functions as both an action planner and a results aggregator, this study adopts the unified dialogue format proposed by LLaVA \cite{liu2023visual}.
As illustrated in Fig. \ref{agent_tooluse}, the system operates through a structured workflow consisting of four key stages: (1) The user provides a task instruction  $\mathrm{X}_{q}$ alongside a contextual image $\mathrm{I}_{q}$. (2) Serving as a central planner and router, the LMM first generates \texttt{thoughts} that analyze \(\mathrm{X}_{q}\) and \(\mathrm{I}_{q}\) to identify the optimal tool from the skill repository for task fulfillment and the corresponding required parameters. It then formulates \texttt{actions} that explicitly specify the API name of the designated tool and its associated parameters, followed by outputting \texttt{value} to inform the user that the selected tool will be invoked for task processing. These three fields collectively constitute the structured command \(\text{X}_{\mathrm{tool\_use}}\).
(3) The designated tool is executed with $\mathrm{I}_{q}$ as input, and its output $\text{X}_{\mathrm{tool\_result}}$ is returned to the LMM. 
(4) Finally, the BAAI Cardiac Agent first generates \texttt{thoughts} that integrate $\text{X}_{\mathrm{tool\_result}}$, $\mathrm{X}_{q}$, and $\mathrm{I}_{q}$ to confirm how to organize a coherent response. It sets \texttt{actions} as an empty list, indicating no additional tools need to be invoked, before synthesizing all information to generate \texttt{value}—the coherent and informed final response $\mathrm{X}_{answer}$ provided to the user.
This interaction process can be represented as:

\begin{align}
\label{equ:agent_inp_out}
    \texttt{User} : \mathrm{I}_q <\texttt{\textbackslash n}> \mathrm{X}_q <\texttt{STOP}> \quad \texttt{LMM} : \text{X}_{\mathrm{tool\_use}} <\texttt{STOP}> \\
    \texttt{User} : \text{X}_{\mathrm{tool\_result}} \mathrm{X}_q <\texttt{STOP}> \quad \texttt{LMM} : \mathrm{X}_{answer} <\texttt{STOP}>
\end{align}

The loss is computed only for the reasoning text (\texttt{thoughts}) related to tool invocation, the tool invocation list (\texttt{actions}), and the final natural language answer (\texttt{value}). Loss for redundant text irrelevant to tool usage is not calculated, ensuring that training resources are concentrated on improving tool usage capabilities.

Based on SAX cine, 2CH cine, 4CH cine, LGE, and Rest MPI, this study generated 32k instruction-tuning data samples, including 8K augmented VQA instructions, as well as 3K samples each for the tasks of SAXCS, 2CHCS, 4CHCS, SAXLGES, CDS, NICMS, RAG, and MRG. The LMM initializes with LLaVA-Med 60K-IM. Lightweight fine-tuning performs on the language model using the Low-Rank Adaptation (LoRA) technique \cite{hu2022lora}, with training conducted for 10 epochs and a batch size set to 8. By training only a small number of low-rank adaptation parameters, the model effectively achieves a favorable balance between computational efficiency and model performance. The LoRA hyperparameter configuration in this experiment is as follows: rank sets to 128, alpha coefficient to 256 and dropout rate to 0.05. To improve training efficiency and reduce memory consumption, the DeepSpeed ZeRO-3 optimization strategy \cite{ren2021zero} adopts during training. It integrates optimizer state sharding, gradient checkpointing technology and FP16 mixed-precision training \cite{micikevicius2017mixed}. In terms of hyperparameter settings, a learning rate of 1e-4 selects for language model fine-tuning, combined with a cosine learning rate scheduler to ensure stable convergence. This relatively high learning rate can effectively facilitate model adaptation to tasks. Meanwhile, the learning rate of the visual projector configures independently to 2e-5. The low learning rate preserves the modality alignment effect achieved in previous training. 

\subsubsection{Tasks and tools of the BAAI Cardiac Agent
}\label{subsubsec232}
The BAAI Cardiac Agent in this study is capable of accessing a diverse range of tools and features scalability to adapt to multi-task processing. As shown in Fig. \ref{tool_task}, we integrate 9 tools covering major representative tasks in the CMR domain.
It should be noted that no additional tools are required for the VQA task, as we utilize LLaVA-Med \cite{li2023llava} as the backbone model, which natively supports the VQA task. Each tool has specialized functional attributes and exhibits exceptional performance in executing specific cardiac magnetic resonance tasks.

For the segmentation tasks (Tasks 1–4), this study develops a general-purpose medical image segmentation architecture independently, as detailed in Section \ref{subsec24}, 
instead of selecting MedSAM2 \cite{ma2025medsam2}, nnUNet \cite{isensee2021nnu}, ResUNet++ \cite{jha2019resunet++}, or DiffUNet \cite{xing2025diff} as the core tools.
This decision is based on two primary considerations. First, accurate segmentation results are the fundamental prerequisite for precisely measuring key cardiac functional parameters. Second, the strong robustness of the segmentation model architecture across multiple CMR sequences is of vital significance for the practical deployment of the BAAI Cardiac Agent. 
This study presents a comparative analysis of segmentation results between the proposed architecture and the aforementioned four SOTA methods (MedSAM2 \cite{ma2025medsam2}, nnUNet \cite{isensee2021nnu}, ResUNet++ \cite{jha2019resunet++}, and DiffUNet \cite{xing2025diff}) in Section \ref{subsec32}.

For the diagnostic tasks (Tasks 5–6), this study designs a two-stage diagnostic framework. First, it performs preliminary screening of NH, IHD, and NICM based on CMR cine data. Second, it incorporates SAX LGE data to enable fine-grained classification of NICM subtypes, including HCM, DCM, RCM, ACM, and myocarditis. Detailed specifications of the diagnostic architecture are provided in Section \ref{subsec25}. To validate the effectiveness of the proposed CMR-based diagnostic method, we conduct fair comparative experiments against SOTA approaches, including the CMR diagnostic method based on VST \cite{wang2024screening}, ViT \cite{dosovitskiy2020image}, and 3D ResNet \cite{qiu2017learning}. As demonstrated in the diagnostic results presented in Section \ref{subsec33}, the reliability of our method is fully verified.

For the generation of CMR medical reports, LVEDWT by the AHA 17-SM can be measured based on segmentation results of SAX cine images, including the anterior wall (AW), anteroseptal wall (ASW), inferoseptal wall (ISW), inferior wall (IW), inferolateral wall (ILW), and anterolateral wall (ALW) at the basal segment; the AW, ASW, ISW, IW, ILW, and ALW at the mid segment; and the AW, SW, IW, and lateral wall (LW) at the apical segment.
Cardiac function analysis of the LV mainly includes the measurement of LVEF, LVEDV, LVESV, SV, CO, and LVM.
LV cardiac apex thickness is measured using segmentation results of 4CH cine images. Based on 4CH cine images, additional parameters can be quantified, including the maximum transverse diameters of the left and right atria measured in the 4CH view, denoted as LAT4CHD and RAT4CHD. 

RAG refers to a technical method that enhances the quality of generated results by integrating the most relevant information acquired from external data sources. In this study, we extract the cardiovascular section from ChatCAD+ \cite{zhao2024chatcad+}, and combine it with clinical guidelines \cite{virani20232023,de2003european,arbelo20232023} related to CVDs as well as heart-related literature retrieved from PubMed, to implement the medical retrieval process.

In the VQA component, the core focus lies on the holistic diagnostic description of CMR data by the BAAI Cardiac Agent. To this end, three categories of structured tasks are designed: (1) Overall CMRSR for clinical cardiac image analysis tasks; (2) Panoramic description of cine sequence data, covering the normality of key anatomical structures and functional indicators including VS, VWM, VSF and VDF, mitral valve MV, TV, and PE, as well as interpretation of Rest MPI, which is primarily employed to evaluate whether the myocardial blood perfusion status conforms to normal physiological standards; (3) Conversational content related to CVDs. 

\subsection{CMR-based expert model for cardiac segmentation
}\label{subsec24}
\begin{figure*}[!t]
	\centering
		\includegraphics[width=1 \textwidth] {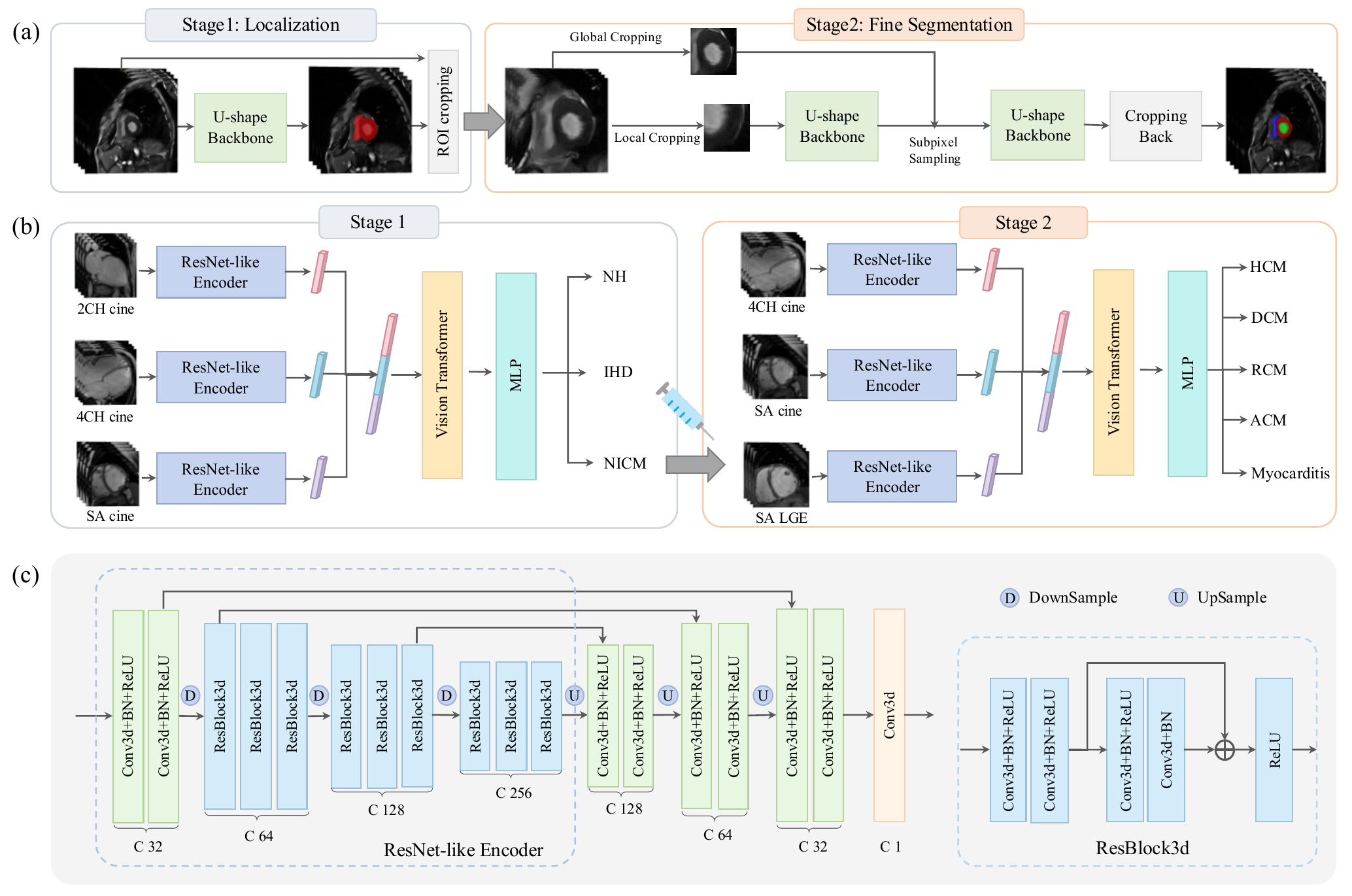} 
	\caption{The architecture of the proposed expert model for cardiac segmentation and CVDs diagnosis is described as follows: (a) It is a two-stage coarse-to-fine cardiac segmentation framework. In the first stage, a U-shaped backbone locates the ROI of the heart. In the second stage, based on the ROI output from the first stage, the framework fuses the global and local features of the heart to generate a refined, high-resolution cardiac segmentation map. (b) It is a two-stage automated screening and diagnostic framework for CVDs. In the first stage, the model leverages cine images (including 2CH, 4CH, and SAX views) to implement tri-class screening, categorizing subjects into three groups: NH, IHD, or NICM. For patients identified as having NICM via screening, the second stage further integrates their retained 4CH cine images, short-axis cine images, and SAX LGE images to output more refined diagnostic results for NICM subtypes. (c) It represents the overall network structure of the U-shaped backbone. 
    } \label{fig:4}
\end{figure*}

The CMR-based expert model for cardiac segmentation adopts the same two-stage coarse-to-fine architecture for 2CH cine, 4CH cine, SAX cine, and SAX LGE data, with the detailed structure illustrated in Fig. \ref{fig:4} (a). The first stage localizes the cardiac region of interest (ROI), and the second stage performs fine-grained segmentation based on the localization results. 

In the first stage of the general cardiac segmentation expert model, coarse cropping is performed on original 3D medical images. The spatial extent of the cropped volume is determined by a randomly sampled center, enabling extraction of a local subvolume containing the target region. Subsequently, the cropped region is resampled to a uniform voxel spacing. During resampling, continuous interpolation is adopted for intensity images to preserve signal continuity, while nearest-neighbor interpolation is utilized for segmentation labels to maintain boundary integrity. After cropping and resampling, the images undergo min-max normalization and are fed into a unified U-shaped backbone, which outputs accurate localization results of the heart.

The second stage of the segmentation expert model is performed based on the ROI localized in the first stage, with its core adopting a dual-path pyramid feature extraction design. This design comprises two cropping branches: global and local. The receptive field of the global cropping region is twice that of the local branch, enabling capture of broader contextual information, while its resolution is only half of the local branch. The local cropping branch focuses on high-resolution detailed features. A random strategy is employed to determine the cropping centers, after which both branches are processed into unified fixed-size image patches (e.g., 32 $\times$ 64 $\times$ 64 for 2CH or 4CH cine, 64 $\times$ 64 $\times$ 64 for SAX cine, and 3 $\times$ 64 $\times$ 64 for SAX LGE). In the feature fusion process, both branches utilize the same U-shaped backbone network configured identically to the first stage. The global branch first performs feature extraction, and then upsamples its feature maps to a spatial size matching the local branch via sub-pixel sampling. The fused features, together with the original locally cropped images, are fed into the U-shaped network of the local branch for deeper and finer-grained feature mining, ultimately outputting accurate cardiac segmentation results. To enhance the model's robustness against anatomical variability and acquisition noise, a 3D data augmentation strategy including random flipping, scaling, translation, elastic deformation, and random field perturbation is implemented during the training process of both stages. 

The U-shaped backbone network is based on the ResUNet \cite{diakogiannis2020resunet} architecture, and its detailed structure is presented in the bottom section of Fig. \ref{fig:4} (c). The encoder of the U-shaped backbone consists of an initial double-convolution block followed by three residual layers, each containing three residual blocks. Downsampling is achieved using stride-2 convolutions, with the number of feature channels progressively increasing. The decoder restores spatial resolution through three upsampling steps; after each upsampling, the features are concatenated with corresponding encoder features via skip connections and fused using convolutional layers. Each residual block comprises two 3 $\times$ 3 $\times$ 3 convolutional layers with Batch Normalization (BN); a ReLU activation follows the first convolution, while the second convolution is not activated before addition with the identity pathway. When channel dimensions or resolutions differ, the identity mapping is aligned through a projection branch using a 1 $\times$ 1 $\times$ 1 convolution with BN. The final segmentation output is generated by a 1 $\times$ 1 $\times$ 1 convolutional layer applied to the backbone features. For LGE data, due to its large slice thickness and limited spatial continuity, the above 3D model architecture designed for cine sequences is adapted into a 2D structure to accommodate the characteristics of LGE imaging.

Both stages of the cardiac segmentation expert model undergo 100 epochs of training, and the weights corresponding to the optimal performance on the validation set are selected for performance inference.
The Adam optimizer is used with an initial learning rate of 5e-4 and a weight decay of 5e-4. A piecewise learning rate schedule with linear warm-up is adopted, where the learning rate is linearly increased from one-third of its initial value to the target value during the first 10 iterations, followed by a decay by a factor of 0.2 at the 5th, 30th, and 60th epochs. 

\subsection{CMR-based expert model for CVDs}\label{subsec25}
CMR technology has emerged as the gold standard for non-invasive assessment of cardiac structure and function \cite{chowdhary2021cardiovascular,parlati2025advancing}, enabling comprehensive visualization of the complete characteristics of myocardial motion and myocardial fibrosis through dynamic cine imaging across multiplanar views and SAX LGE imaging \cite{rogers2025identification}. The CMR-based expert model for CVDs adopts a two-stage architecture, as illustrated in Fig. \ref{fig:4} (b). Owing to the significant differences in the motion characteristics of cardiac contraction among NH, IHD, and NICM, the first stage exclusively utilizes multi-view cine sequences from CMR SAX, long-axis 2CH, and long-axis 4CH views. By integrating motion features extracted from multi-view cine data via deep learning algorithms, this stage achieves automated differentiation among three major categories: NH, IHD, and NICM, thereby avoiding the potential risks associated with contrast agent injection. However, given the complexity of diagnosing NICM, reliable assessment typically requires LGE technology. Thus, the second stage incorporates SAX LGE sequences to improve the diagnostic accuracy of fine-grained classification for NICM. Myocardial fibrosis manifests distinctively across different NICM \cite{cojan2021non,giordano2022myocardial}.

In the first stage of the CVDs expert model, the input comprises 2CH, 4CH, and SAX cine sequences, each accompanied by a corresponding segmentation mask. The preprocessing pipeline includes three steps: z-axis cropping, voxel spacing resampling, and xy-plane cropping. Since z-axis cropping is related to cardiac cycle phases, the central three phases are retained for 2CH and 4CH sequences, and the central nine phases for SAX sequences. All data are then resampled to a uniform voxel spacing, defined as the mode spacing of the dataset. The ROI is extracted using the segmentation mask, and images are cropped to a fixed size of 200 $\times$ 200 in the xy-plane centered at the ROI. After this initial preprocessing, a second cropping step is applied: 2CH and 4CH sequences are cropped to 80 $\times$ 192 $\times$ 192, and SAX sequences to 288 $\times$ 144 $\times$ 144. Intensity normalization is then performed, followed by probabilistic color augmentation or noise perturbation. During training, additional sample-wise 3D data augmentations are applied, including random 3D rotation, scaling, translation, flipping, and interpolation, to improve model robustness. 

To effectively capture modality-specific characteristics from high-dimensional 3D data, separate ResNet-like \cite{he2016deep,hara2017learning} encoders are designed for the 2CH, 4CH, and SAX modalities, with the detailed structure illustrated in the lower half of Fig. \ref{fig:4} (c). Each ResNet-like encoder in the CVDs diagnosis expert model shares the same architecture, consisting of an initial double-convolution block, three residual layers (each with three basic blocks), and a global average pooling layer. Each basic block is a 3D residual unit composed of two 3 $\times$ 3 $\times$ 3 convolutions with BN; the first convolution is followed by a ReLU activation, whereas the second is not activated before addition with the identity mapping. When channel dimensions or resolutions differ, a projection branch with 1 $\times$ 1 $\times$ 1 convolution and BN (optionally with downsampling) is used for alignment. The modality-specific features are concatenated and fused using a Vision Transformer (ViT) \cite{dosovitskiy2020image}, and the transformer \cite{vaswani2017attention} outputs are mapped to classification logits using a multilayer perceptron (MLP) \cite{popescu2009multilayer}.

The second stage of the CVDs expert model builds upon the first stage by replacing the 2CH cine input with the LGE modality. The preprocessing pipeline for LGE follows that of cine data: nine central slices are retained along the z-axis, and images are cropped to 200 $\times$ 200 in the xy-plane, followed by a second cropping step to 9 $\times$ 144 $\times$ 144. Unlike cine encoders, the LGE encoder does not compress the z-axis dimension to preserve spatial information relevant to scar localization.

The two-stage CVDs diagnosis model is trained for 60 epochs using the Adam optimizer with an initial learning rate of 5e-4 and a weight decay of 5e-4. A piecewise learning rate schedule with linear warm-up is employed, where the learning rate is linearly increased from one-third of its initial value to the target value during the first 10 iterations and reduced by a factor of 0.2 at the 5th and 20th epochs.

\bibliographystyle{unsrt}  
\bibliography{references}


\section*{Data availability}
IRB approval was obtained from all participating institutions for imaging and data collection: Beijing Anzhen Hospital, China (2025216x). The need for informed consent was waived by the respective ethics committees and institutions. The benchmark test dataset for multi-modal large model evaluation has been publicly released, with access available at \href{https://huggingface.co/datasets/TaipingQu/CMRAgentEvalSet}{https://huggingface.co/datasets/TaipingQu/CMRAgentEvalSet}. We have uploaded all pixel-level multi-sequence CMR annotations to \href{https://huggingface.co/datasets/TaipingQu/CMR-MULTI}{https://huggingface.co/datasets/TaipingQu/CMR-MULTI}. No other publicly available datasets were used in this study. The deidentified data can be shared only for noncommercial academic purposes and will require a formal material transfer agreement and a data use agreement. Requests should be submitted by emailing the corresponding authors. All requests will be evaluated based on institutional policies to determine whether the data requested are subject to intellectual property or patient privacy obligations.
\section*{Code availability}
An open-source version of the code base is available on GitHub at \href{https://github.com/plantain-herb/Cardiac-Agent}{https://github.com/plantain-herb/Cardiac-Agent} with no restrictions.
\section*{Acknowledgements}
This work was supported by the grants from the National Key R\&D Program of China (2022YFE0209800), received by L. Xu, the National Natural Science Foundation of China (82271986, U1908211), received by L. Xu, the Beijing Natural Science Foundation (7244326 \&L246062), received by HK. Zhang and L. Xu.
\section*{Author contributions}
T. Qu, H.K. Zhang, L. Xu and H.G. Zhang conceptualized the study collaboratively. T. Qu, L. Zhang, C. Zhao, M. Zou, P. Zhao, H. Liu, and Z. Su were responsible for cardiac magnetic resonance (CMR) data preprocessing, large model algorithm development and training, expert model training, as well as comprehensive analysis and interpretation of model performance. K. Bo, X. Jin, H.K. Zhang, Z. Zhou, N. Zhang, H. Wang, K. Jiang, Y. Du, and M. Wang primarily undertook the collection, annotation and quality control of clinical CMR data, and completed the clinical efficacy evaluation of the model. R. Yan coordinated the acquisition of external validation CMR data. Z. Wang and T. Huang provided overall project coordination, technical platform support and funding guarantee for the entire study. L. Xu offered professional clinical guidance, coordinated multi-center clinical resources, and conducted rigorous review of all clinical-related content in the study. L. Zhang and C. Zhao contributed equally and are co-second authors. T. Qu, H.K. Zhang, L. Zhang, and C. Zhao drafted the initial manuscript and prepared the key figures. L. Xu and H.G. Zhang were responsible for the overall study design and academic guidance, manuscript finalization and revision, as well as all communication work with the journal. All authors provided critical feedback on the study design and manuscript, and ensured the scientific integrity of this work.
\section*{Competing interests}
The authors declare no competing interests.
\begin{appendices}
\renewcommand{\figurename}{Fig.}
\renewcommand{\tablename}{Table}
\renewcommand{\thefigure}{D\arabic{figure}}
\renewcommand{\thetable}{D\arabic{table}}
\setcounter{figure}{0}
\setcounter{table}{0}
\section{Study Cohort Statistics}\label{secA1}

\subsection{CMR Acquisition}\label{secA11}
All patients underwent CMR scanning with a 32-channel phased array coil under respiratory navigation and electrocardiographic gating in the supine position. The scanning equipment was three 3.0 T CMR whole-body scanner: Achieva, Philips, Netherlands, Holland; Discovery MR750w, GE Healthcare, USA; Siemens Healthineers AG, Erlangen, Germany. The cine images were acquired using retrospective electrocardiogram (ECG) gating, and the standardized imaging protocol included steady-state free precession (SSFP) breath-holding cine images obtained at the end of inspiration. These images cover short-axis (SAX) and long-axis views; the latter specifically includes two-chamber (2CH) and four-chamber (4CH) views. The protocol also incorporates late gadolinium enhancement (LGE) images \cite{kramer2020standardized}. Briefly, SAX cardiac images were acquired to cover the entire left ventricle (LV), from the mitral valve annulus to the level of the LV apex, with 25–30 phases acquired per cardiac cycle and a slice thickness of 8 mm. For long-axis views, images were acquired with a slice thickness of 5 mm, encompassing 2CH, 4CH, and three-chamber (3CH) views. LGE images were acquired 10–15 minutes after the intravenous injection of gadolinium-diethylenetriamine pentaacetic acid (Gd-DTPA; dose: 0.2 mmol/kg; Bayer Pharma AG, Berlin, Germany), using a prospectively ECG-gated breath-hold phase-sensitive inversion recovery segmented gradient-echo sequence.

\subsection{Study Cohort}\label{secA12}
We retrospectively collect CMR scan sequences from 2,134 patients (1,558 male and 576 female) at Beijing Anzhen Hospital, Capital Medical University (Beijing, China), with corresponding CMR imaging reports available for all patients. To ensure model reliability, the enrollment period at this center is extended to January 1, 2016, to January 1, 2024, for rare cardiovascular diseases (RCM and ACM). Additionally, we collect CMR scan sequences from 279 patients at the First Affiliated Hospital of Xinxiang Medical University (Henan, China), covering the period from October 1, 2023, to October 15, 2025. As a national-level diagnosis and treatment center for cardiovascular diseases, Beijing Anzhen Hospital provides primary national representative sample support for the dataset. The baseline CMR scan of each examination in both datasets includes SAX cine sequences, 4CH cine sequences, 2CH cine sequences, Rest MPI, and SAX LGE sequences.

\begin{table*}[htbp]
\centering
\renewcommand{\arraystretch}{1.2}
\setlength{\tabcolsep}{3.5pt}
\small
\rowcolors{3}{gray!15}{white} 
\begin{tabular}{@{}lcccccccc@{}}
\toprule
& \multicolumn{4}{c}{\textbf{Internal Dataset}} & \multicolumn{4}{c}{\textbf{External Dataset}} \\  
\cmidrule(lr){2-5} \cmidrule(lr){6-9} 
& \textbf{N} & \textbf{Male} & \textbf{Female} & \textbf{Age} 
& \textbf{N} & \textbf{Male} & \textbf{Female} & \textbf{Age} \\
\midrule
Total & 2134 & 1558 (73\%) & 576 (27\%) & 47±20 & 279 & 171 (61\%) & 108 (39\%) & 49±17 \\
NH & 322 & 190 (59\%) & 132 (41\%) & 22±17 & 67 & 26 (39\%) & 41 (61\%) & 43±21 \\
IHD & 805 & 676 (84\%) & 129 (16\%) & 57±12 & 49 & 37 (76\%) & 12 (24\%) & 56±12 \\
NICM & 830 & 564 (68\%) & 266 (32\%) & 48±18 & 159 & 107 (67\%) & 52 (33\%) & 49±16 \\
HCM & 402 & 269 (67\%) & 133 (33\%) & 51±16 & 43 & 32 (74\%) & 11 (26\%) & 51±14 \\
DCM & 136 & 101 (74\%) & 35 (26\%) & 49±16 & 91 & 62 (68\%) & 29 (32\%) & 52±15 \\
RCM & 95 & 59 (62\%) & 36 (38\%) & 54±20 & 6 & 4 (67\%) & 2 (33\%) & 44±12 \\
ACM & 66 & 46 (69\%) & 20 (31\%) & 43±17 & 8 & 5 (63\%) & 3 (37\%) & 50±22 \\
Myocarditis & 131 & 85 (65\%) & 46 (35\%) & 23±17 & 11 & 4 (36\%) & 7 (64\%) & 26±17 \\
Others & 177 & 115 (65\%) & 62 (35\%) & 46±19 & 4 & 1 (25\%) & 3 (75\%) & 63±16 \\
\bottomrule
\end{tabular}

\caption{Characteristics of the internal and external validation datasets (N: Number of subjects, Age in years: mean±SD (range))}
\label{demographics}
\end{table*}

\section{Model Evaluation Metrics}\label{secAcal}
To assess the CMR segmentation performance, we employ three primary evaluation metrics. The Dice Similarity Coefficient (DSC), calculated as 
 \begin{equation} \label{equ:DSC}
 DSC=\frac{2|\mathbf V_{X}\cap \mathbf V_{Y}|}{|\mathbf V_{X}|+|\mathbf V_{Y}|},
\end{equation}
quantifies the volumetric overlap between the predicted segmentation $\mathbf V_{X}$ and the reference ground truth $\mathbf V_{Y}$. 
Let $\mathbf S_{X}$ and $\mathbf S_{Y}$ represent the voxel collections lying on the boundaries of the predicted segmentation and ground truth, respectively. The Hausdorff Distance (HD) is defined as
 \begin{equation} \label{equ:HD}
 HD=max\left\{{\max_{x\in \mathbf S_{X}}{\min_{y\in \mathbf S_{Y}}}||x-y||,{\max_{y\in \mathbf S_{Y}}{\min_{x\in \mathbf S_{X}}}||y-x||}}\right\},
\end{equation}
where $||$.$||$ denotes the Euclidean distance. This metric is used to evaluate boundary alignment.
Additionally, we use the Average Surface Distance (ASD), a distance-centric metric computed as
\begin{equation} \label{equ:ASD}
 ASD=\frac{1}{|\mathbf S_{Y}|}(\sum_{y\in \mathbf S_{Y}}\min_{x\in \mathbf S_{X}}||y-x||),
\end{equation}
where $|$.$|$ denotes the cardinality of a set. For distance-based metrics, smaller values correspond to superior segmentation outcomes.

To comprehensively quantify the diagnostic performance of the model for CVDs, this study adopts accuracy, sensitivity, specificity, F$_1$-score, and AUC as the core evaluation metrics. All metrics are calculated based on the confusion matrix constructed from diagnostic results, with their specific definitions, formulas, and implications presented below. The core parameters of the confusion matrix are defined as follows: true positive (TP), true negative (TN), false positive (FP), and false negative (FN). The core calculation formulas are as follows.

\begin{equation} \label{equ:acc}
Accuracy = \frac{TP + TN}{TP + TN + FP + FN}
\end{equation}
\begin{equation} \label{equ:acc}
Sensitivity = \frac{TP}{TP + FN}
\end{equation}
\begin{equation} \label{equ:acc}
Specificity = \frac{TN}{TN + FP}
\end{equation}
\begin{equation} \label{equ:acc}
Precision = \frac{TP}{TP + FP}
\end{equation}
\begin{equation} \label{equ:acc}
F_1\text{-score} = 2 \times \frac{Precision \times Sensitivity}{Precision + Sensitivity}
\end{equation}
\section{CMR Data Annotation}\label{secA4}
CMR  annotations  were  performed  independently  by two  radiologists(H.K.Z and N.Z) with  no less than 10  years of clinical experience in CMR diagnosis, using the open-source software ITK-SNAP (www.itksnap.org)  for  precise  contour  delineation.  The  annotation  scope  for  each sequence is as follows: the SAX cine sequence includes the LV cavity, LV myocardium, and right ventricle (RV); the 4CH cine sequence covers the LV cavity, LV myocardium, RV cavity, RV myocardium, left atrium (LA), and right atrium (RA); the 2CH cine sequence involves the LV cavity and LV myocardium; and the SAX LGE sequence encompasses the LV cavity, LV myocardium, and LGE regions. After annotation completion, a senior radiologist(L.X, more than 20 years cardiac MRI experience) verifies and validates the annotation quality of all data to ensure accuracy and consistency.

\section{CMR Report Scoring Criteria
}\label{secA5}
The CMR report scoring system has a total of 100 points and comprises two modules: clinical diagnostic accuracy (70 points) and technical performance (30 points). The core of clinical diagnostic accuracy is assessing the consistency between model output and the clinical gold standard. Specifically, clinical diagnosis (20 points) evaluates the clarity and correctness of cardiac disease diagnosis; core structural quantification (15 points) involves score deductions based on the errors between automatically measured values including end-diastolic volume, end-systolic volume and ejection fraction and the clinical report text. Full marks are given when the error is less than 5\%, with 3 points deducted for an error ranging from 5\% to 10\%, 5 points for 10\% to 20\%, and 7 points for more than 20\%. Wall assessment (10 points), myocardial viability detection via LGE (15 points) and other feature assessment (10 points) respectively evaluate the integrity of key information detection for the corresponding items, with 1 point deducted for each missing key piece of information. The other features cover indicators such as cine valve status, cardiac size, ventricular wall motion, pericardial effusion and myocardial perfusion assessment. For the technical performance module, only report completeness and structuring (30 points) is assessed, which judges whether the overall CMR report complies with clinical norms. Two points are deducted for each piece of redundant information and also for each missing key clinical indicator. In addition, severe score deductions ranging from 0 to -10 points are imposed for risky behaviors such as model hallucinations, including unfounded speculation and fabrication of non-existent information. The deductions are graded in accordance with major factual hallucinations, mild factual hallucinations or over-interpretation, and logically conflicting hallucinations.

\begin{table}[htbp]
  \centering
  \small 
  \begin{tabular}{c>{\columncolor{white}}lccc}
    \toprule
    Metric & Method & LV Myo & LV Cavity & RV Cavity\\
    \midrule
    \rowcolor{gray!15}
    \multicolumn{1}{c}{\cellcolor{white}}& ResUNet++ & 91.29±0.03 & 85.14±0.04 & 77.64±0.08 \\
    & nnUNet & 88.19±0.04 & 92.44±0.03 & 81.63 ± 0.05 \\
    \rowcolor{gray!15}
    \multicolumn{1}{c}{\cellcolor{white}DSC} & DiffUNet & 83.37±0.16 & 80.34±0.11 & 67.88±0.12 \\
    & MedSAM2 & 78.60±0.28 & 82.95±0.31 & 81.02±0.07 \\
    \rowcolor{gray!15}
    \multicolumn{1}{c}{\cellcolor{white}}& Ours & \textbf{92.44±0.03} & \textbf{96.19±0.02} & \textbf{82.00±0.05} \\
    \midrule
    & ResUNet++ & 9.67±5.99 & 9.01±3.47 & 31.07±24.36 \\
    \rowcolor{gray!15}
    \multicolumn{1}{c}{\cellcolor{white}}& nnUNet & 8.04±2.91 & 7.13±2.32 & 25.87±29.31 \\
    HD & DiffUNet & 16.44±8.05 & 184.23±53.80 & 76.23±62.99 \\
    \rowcolor{gray!15}
    \multicolumn{1}{c}{\cellcolor{white}}& MedSAM2 & 21.24±6.70 & 18.45±7.19 & 18.48±7.51 \\
    & Ours & \textbf{6.49±2.01} & \textbf{5.83±1.48} & \textbf{12.41±5.45} \\
    \midrule
    \rowcolor{gray!15}
    \multicolumn{1}{c}{\cellcolor{white}}& ResUNet++ & 0.67±0.89 & 0.38±0.13 & 1.75±1.39 \\
    & nnUNet &  \textbf{0.26±0.09} & 0.29±0.14 & 1.71±2.33 \\
    \rowcolor{gray!15}
    \multicolumn{1}{c}{\cellcolor{white}ASD} & DiffUNet & 1.41±1.55 & 2.28±2.08 & 4.26±4.99 \\
    & MedSAM2 & 4.87±1.61 & 0.59±0.29 & 1.90±1.11 \\
    \rowcolor{gray!15}
    \multicolumn{1}{c}{\cellcolor{white}}& Ours & 0.28±0.09 & \textbf{0.14±0.09} & \textbf{1.51±0.79} \\
    \bottomrule
  \end{tabular}
  \caption{A comparative study between the BAAI Cardiac Agent segmentation expert model and SOTA segmentation networks on the SAX cine dataset. We report DSC (\%), HD, and ASD as mean ± standard deviation.}
  \label{segressax}
\end{table}

\begin{table}[htbp]
  \centering
  \small 
  \begin{tabular}{clcc}
    \toprule
    Metric & Method & LV Myo & LV Cavity\\
    \midrule
    \rowcolor{gray!15}
    \multicolumn{1}{c}{\cellcolor{white}}& ResUNet++ & 83.39±0.05 & 90.56±0.03 \\
    & nnUNet & 85.49±0.03 & 90.79±0.04 \\
    \rowcolor{gray!15}
    \multicolumn{1}{c}{\cellcolor{white}DSC} & DiffUNet & 78.92±0.09 & 87.17±0.07 \\
    & MedSAM2 & \textbf{87.17±0.04} & 68.61±0.40 \\
    \rowcolor{gray!15}
    \multicolumn{1}{c}{\cellcolor{white}}
    & Ours & 84.95±0.05 & \textbf{92.54±0.03} \\
    \midrule
    & ResUNet++ & 13.84±10.82 & 10.71±4.28 \\
    \rowcolor{gray!15}
    \multicolumn{1}{c}{\cellcolor{white}}
    & nnUNet & 29.25±51.57 & 30.51±50.16 \\
    HD & DiffUNet & 189.56±64.42 & 45.15±58.17 \\
    \rowcolor{gray!15}
    \multicolumn{1}{c}{\cellcolor{white}}
    & MedSAM2 & \textbf{7.42±2.85} & 11.02±7.12 \\
    & Ours & 8.43±3.57 & \textbf{7.31±2.75} \\
    \midrule
    \rowcolor{gray!15}
    \multicolumn{1}{c}{\cellcolor{white}}
    & ResUNet++ & 0.55±0.35 & 0.73±0.51 \\
    & nnUNet & 2.14±6.54 & 3.30±9.56 \\
    \rowcolor{gray!15}
    \multicolumn{1}{c}{\cellcolor{white}ASD}
    & DiffUNet & 8.05±10.25 & 2.88±6.25 \\
    & MedSAM2 & \textbf{0.18±0.11} & 0.32±0.27 \\
    \rowcolor{gray!15}
    \multicolumn{1}{c}{\cellcolor{white}}
    & Ours & 0.29±0.17 & \textbf{0.25±0.18} \\
    \bottomrule
  \end{tabular}
  \caption{A comparative analysis of the BAAI Cardiac Agent segmentation expert model against current SOTA segmentation networks using the 2CH cine dataset. Results are presented as mean ± standard deviation for DSC (\%), HD, and ASD.}
  \label{segres2ch}
\end{table}

\begin{table}[htbp]
  \centering
  \small 
  \setlength{\tabcolsep}{3pt}
  \begin{tabular}{clcccccc}
    \toprule
    Metric & Method & LV Myo & LV Cavity & RV Myo & RV Cavity & LA Cavity & RA Cavity\\
    \midrule
    \rowcolor{gray!15}
    \multicolumn{1}{c}{\cellcolor{white}}
    & ResUNet++ & 83.75±0.04 & 92.94±0.03 & 62.17±0.14 & 88.43±0.04 & 91.37±0.04 & 87.82±0.06 \\
    & nnUNet & 82.34±0.06 & 92.23±0.03 & 57.64±0.14 & 85.39±0.04 & 86.69±0.11 & 85.37±0.09 \\
    \rowcolor{gray!15}
    \multicolumn{1}{c}{\cellcolor{white}DSC}
    & DiffUNet & 78.31±0.06 & 88.02±0.07 & 54.81±0.15 & 79.29±0.11 & 86.32±0.10 & 77.51±0.17\\
    & MedSAM2 & \textbf{86.57±0.03} & 62.87±0.44 & \textbf{70.54±0.07} & \textbf{90.26±0.02} & 91.61±0.05 & \textbf{90.54±0.04}\\
    \rowcolor{gray!15}
    \multicolumn{1}{c}{\cellcolor{white}}
    & Ours & 86.12±0.05 & \textbf{94.25±0.02} & 68.77±0.09 & 89.93±0.03 & \textbf{92.23±0.04} & 90.22±0.05 \\
    \midrule
    & ResUNet++ & 15.81±22.46 & 16.43±25.82 & 10.61±4.41 & 10.45±4.79 & 7.27±4.52 & 21.81±33.01\\
    \rowcolor{gray!15}
    \multicolumn{1}{c}{\cellcolor{white}}
    & nnUNet & 13.41±7.64 &  9.89±4.76 & 14.22±9.70 & 12.21±6.58 & 12.32±6.86 & 10.96±6.91\\
    HD & DiffUNet & 114.55±85.95 & 70.62±81.97 & 67.94±54.54 &  45.69±51.56 & 88.73±80.80 & 50.46±58.70\\
    \rowcolor{gray!15}
    \multicolumn{1}{c}{\cellcolor{white}}
    & MedSAM2 & \textbf{7.08±3.06} & 10.76±7.35 & \textbf{9.22±4.56} & 9.18±3.72 & \textbf{5.99±2.24} & \textbf{5.43±2.14}\\
    & Ours & 7.34±3.59 & \textbf{6.22±2.48} & 9.63±4.78 & \textbf{8.25±3.72} & 6.66±3.10 & 7.00±3.10\\
    \midrule
    \rowcolor{gray!15}
    \multicolumn{1}{c}{\cellcolor{white}}
    & ResUNet++ & 0.61±0.94 & 1.25±2.46 & 0.26±0.18 & 0.47±0.34 & 0.35±0.29 & 2.74±6.64\\
    & nnUNet & 0.37±0.25 & 0.49±0.41 & 0.57±0.75 & 0.89±1.89 & 0.93±1.03 & 0.79±1.10\\
    \rowcolor{gray!15}
    \multicolumn{1}{c}{\cellcolor{white}ASD}
    & DiffUNet & 8.91±11.34 & 6.73±8.00 & 9.41±12.27 & 7.90±8.83 & 11.12±14.44 & 5.48±7.72\\
    & MedSAM2 & \textbf{0.18±0.09} & 0.31±0.24 & \textbf{0.19±0.10} & \textbf{0.21±0.09} & 0.36±0.35 & 0.30±0.18\\
    \rowcolor{gray!15}
    \multicolumn{1}{c}{\cellcolor{white}}
    & Ours & 0.29±0.14 & \textbf{0.20±0.14} & 0.32±0.27 & 0.26±0.13 & \textbf{0.25±0.27} & \textbf{0.23±0.24}\\
    \bottomrule
  \end{tabular}
  \caption{A comparative study of the BAAI Cardiac Agent segmentation expert model versus SOTA segmentation networks on the 4CH cine dataset. We report DSC (\%), HD, and ASD in the format of mean ± standard deviation.}
  \label{segres4ch}
\end{table}

\begin{table}[htbp]
  \centering
  \small 
  \begin{tabular}{clccc}
    \toprule
    Metric & Method & LV Myo & LV Cavity & LGE \\
    \midrule
    \rowcolor{gray!15}
    \multicolumn{1}{c}{\cellcolor{white}}
    & ResUNet++ & 66.40±0.12 & 81.72±0.09 & 26.13±0.26 \\
    & nnUNet & \textbf{74.12±0.10} & 87.36±0.06 & 60.19±0.21 \\
    \rowcolor{gray!15}
    \multicolumn{1}{c}{\cellcolor{white}DSC}
    & DiffUNet & 53.09±0.15 & 58.85 ± 0.25 & 29.88±0.19 \\
    & MedSAM2 & 70.07±0.27 & 72.27±0.34 & \textbf{65.37±0.23} \\
    \rowcolor{gray!15}
    \multicolumn{1}{c}{\cellcolor{white}}
    & Ours & 73.67±0.08 & \textbf{88.60±0.05} & 60.83±0.13 \\
    \midrule
    & ResUNet++ & 13.13±5.60 & 9.88±4.35 & 26.68±14.25 \\
    \rowcolor{gray!15}
    \multicolumn{1}{c}{\cellcolor{white}}
    & nnUNet & 12.31±6.07 & 9.03±5.14 & 16.71±9.12 \\
    HD & DiffUNet & 194.77±55.08 & 157.69±79.11 & 185.30±44.29 \\
    \rowcolor{gray!15}
    \multicolumn{1}{c}{\cellcolor{white}}
    & MedSAM2 & 12.78±5.80 & 9.05±4.09 & \textbf{14.90±9.82} \\
    & Ours & \textbf{11.20±5.14} & \textbf{7.89±3.60} & 24.17±13.85 \\
    \midrule
    \rowcolor{gray!15}
    \multicolumn{1}{c}{\cellcolor{white}}
    & ResUNet++ & 0.60±0.35 & 1.15±0.65 & 2.29±4.64 \\
    & nnUNet & \textbf{0.52±0.24} & 0.59±0.42 & \textbf{0.78±0.58} \\
    \rowcolor{gray!15}
    \multicolumn{1}{c}{\cellcolor{white}ASD}
    & DiffUNet & 48.98±59.52 & 48.62±44.86 & 44.15±38.37 \\
    & MedSAM2 & 0.56±0.39 & 0.80±0.84 & 2.18±4.17 \\
    \rowcolor{gray!15}
    \multicolumn{1}{c}{\cellcolor{white}}
    & Ours & 0.61±0.27 & \textbf{0.54±0.48} & 2.42±2.43 \\
    \bottomrule
  \end{tabular}
  \caption{A comparative analysis of the BAAI Cardiac Agent segmentation expert model and leading segmentation networks evaluated on the SAX LGE dataset. Performance metrics (DSC (\%), HD, and ASD) are reported as mean ± standard deviation.}
  \label{segreslge}
\end{table}

\begin{figure*}[!t]
	\centering
		\includegraphics[width=1 \textwidth] {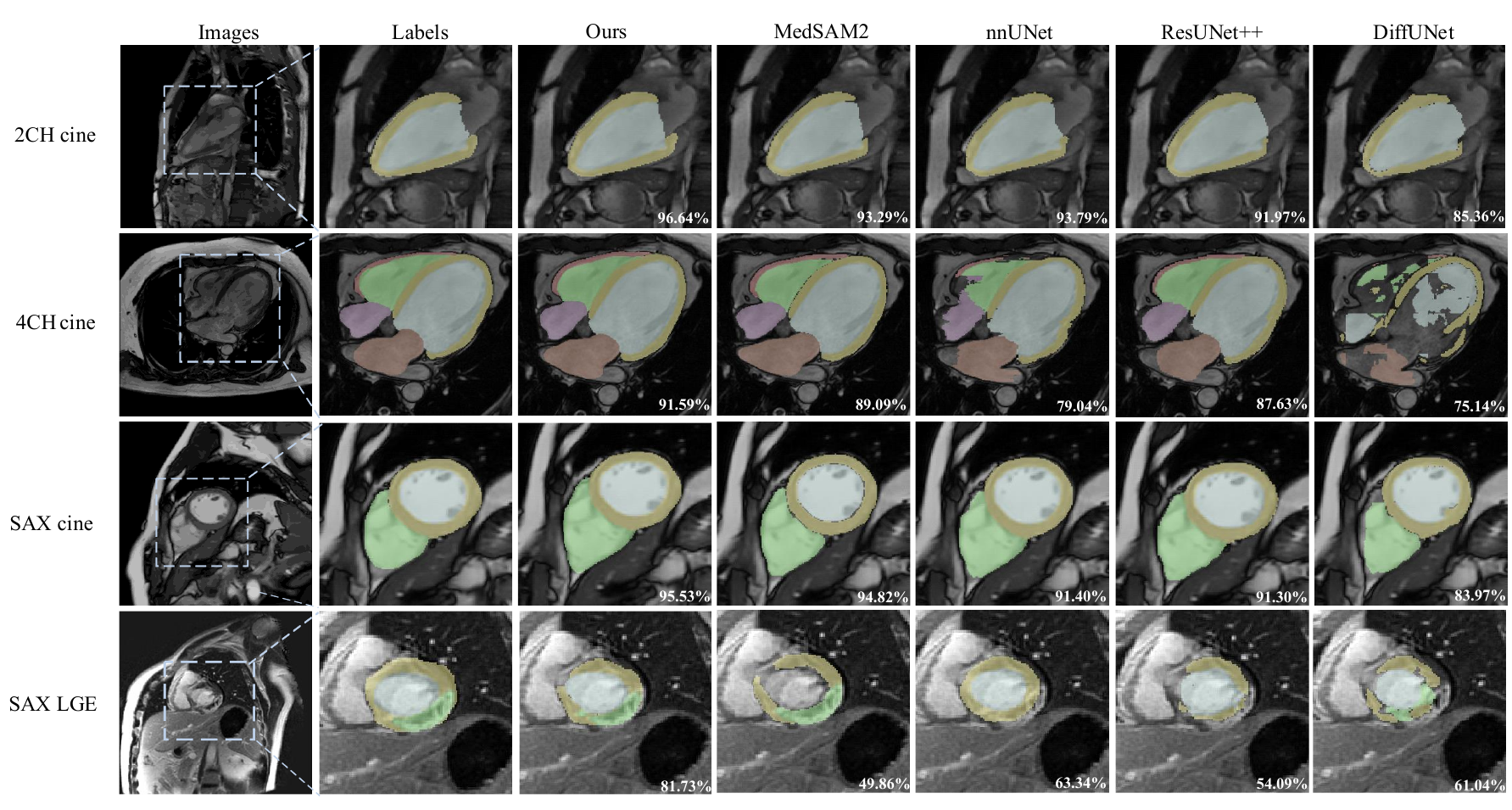} 
	\caption{Qualitative comparison examples of segmentation results between the segmentation expert model we proposed and existing SOTA methods. From top to bottom, four image examples for each of 2CH cine, 4CH cine, SAX cine, and SAX LGE are presented in sequence, with the labeled values indicating DSC results. (Best viewed in color)
    } \label{s1}
\end{figure*}

\begin{figure*}[!t]
	\centering
	\includegraphics[width=1 \textwidth] {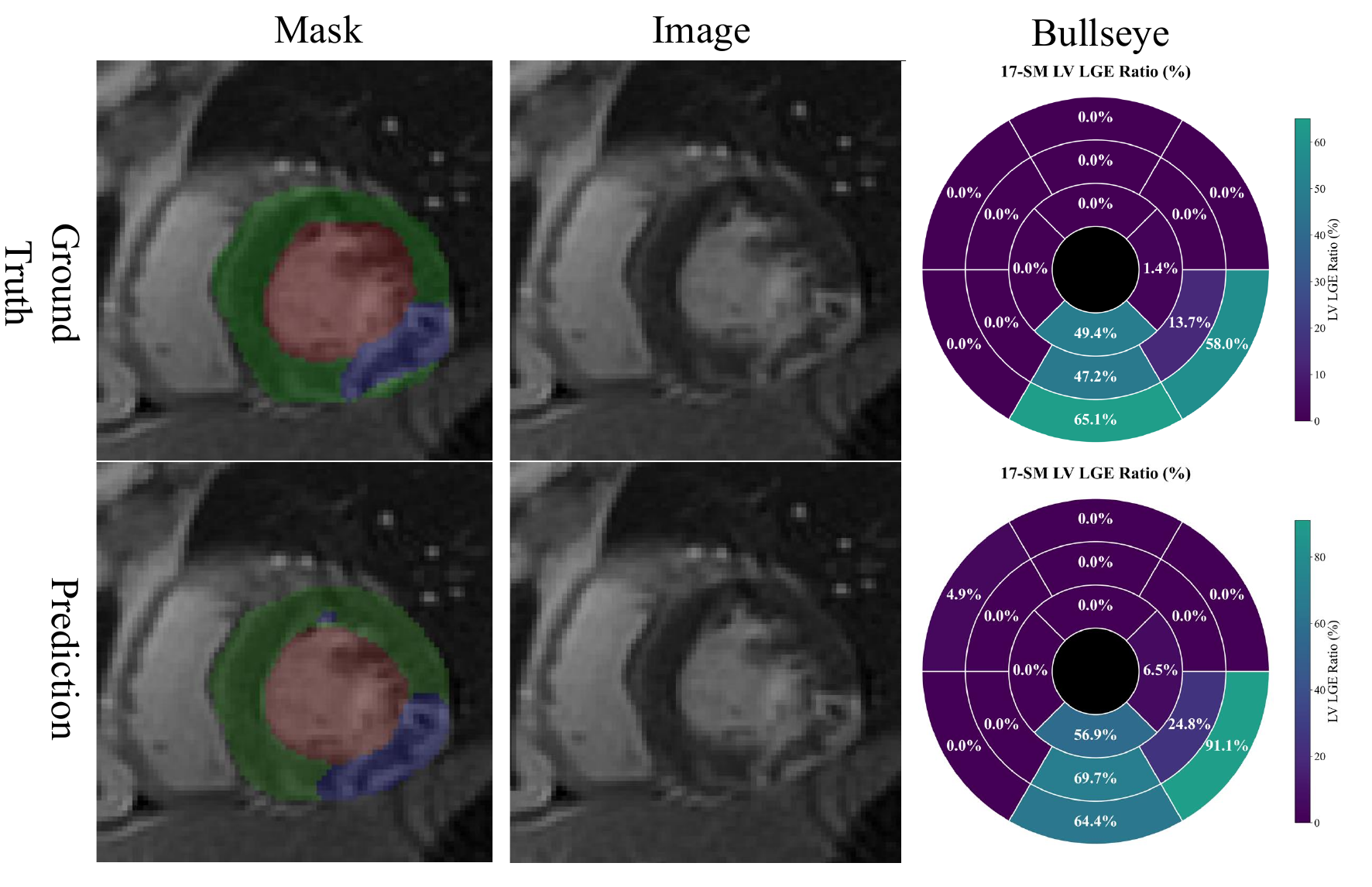} 
	\caption{Cumulative LGE degree in the LV 17-SM: BAAI Cardiac Agent vs physician annotations. (Best viewed in color)
    } \label{LGEBullseye}
\end{figure*}

\begin{table*}[htbp]
  \centering
  \renewcommand{\arraystretch}{1.2}
  \begin{adjustbox}{width=\textwidth,center}
      \begin{tabular}{@{}lcccc@{}}
        \toprule
        \rowcolor{gray!15}
        \multicolumn{5}{c}
        {\textbf{AUC}} \\
        \multicolumn{5}{c}{\textbf{Internal Test Set}} \\
        \midrule
        & \textbf{Ours} & \textbf{VST} & \textbf{ViT} & \textbf{ResNet} \\
        \midrule
        \rowcolor{gray!15}
        NH & \textbf{0.980 (0.956-0.995)} & 0.962 (0.944-0.977) & 0.978 (0.968-0.996) & \textbf{0.980 (954-0.995)} \\
        IHD & \textbf{0.938 (0.913-0.960)} & 0.927 (0.901-0.950) & 0.917 (0.882-0.948) & 0.930 (0.904-0.953) \\
        \rowcolor{gray!15}
        NICM & \textbf{0.960 (0.942-0.974)} & 0.942 (0.920-0.961) & 0.942 (0.920-0.963) & 0.934 (0.908-0.957) \\
        \midrule[0.5pt]
        HCM & \textbf{0.975 (0.948-0.992)} & 0.970 (0.953-0.993) & 0.965 (0.941-0.982) & 0.969 (0.957-0.995) \\
        DCM & 0.971 (0.945-0.991) & 0.969 (0.942-0.990) & \textbf{0.977 (0.953-0.995)} & 0.971 (0.944-0.990) \\
        \rowcolor{gray!15}
        RCM & 0.961 (0.925-0.988) & 0.962 (0.927-0.989) & 0.957 (0.916-0.989) & \textbf{0.975 (0.918-0.992)} \\
        ACM & 0.958 (0.912-0.992) & \textbf{0.959 (0.822-0.990)} & 0.951 (0.898-0.990) & 0.945 (0.900-0.983) \\
        \rowcolor{gray!15}
        Myocarditis & \textbf{0.938 (0.868-0.989)} & 0.929 (0.882-0.974) & 0.927 (0.869-0.986) & 0.841 (0.785-0.887) \\
        \midrule[1.5pt]
        
        \multicolumn{5}{c}{\textbf{External Validation Set}} \\
        \midrule
        & \textbf{Ours} & \textbf{VST} & \textbf{ViT} & \textbf{ResNet} \\
        \midrule
        \rowcolor{gray!15}
        NH & \textbf{0.933 (0.903-0.960)} & 0.920 (0.890-0.947) & 0.891 (0.849-0.929) & 0.916 (0.899-0.958) \\
        IHD & 0.827 (0.767-0.885) & \textbf{0.830 (0.746-0.889)} & 0.782 (0.693-0.861) & 0.805 (0.742-0.864) \\
        \rowcolor{gray!15}
        NICM & \textbf{0.817 (0.770-0.862)} & 0.802 (0.774-0.868) & 0.773 (0.713-0.825) & 0.794 (0.740-0.849) \\
        \midrule[0.5pt]
        HCM & \textbf{0.907 (0.850-0.953)} & 0.881 (0.833-0.938) & 0.884 (0.838-0.941) & 0.903 (0.866-0.951) \\
        DCM & \textbf{0.867 (0.809-0.921)} & 0.859 (0.832-0.915) & 0.822 (0.767-0.891) & 0.830 (0.806-0.906) \\
        \rowcolor{gray!15}
        RCM & 0.877 (0.793-0.939) & 0.873 (0.789-0.950) & \textbf{0.915 (0.843-0.964)} & 0.880 (0.776-0.959) \\
        ACM & 0.902 (0.783-0.977) & 0.911 (0.817-0.969) & \textbf{0.921 (0.756-0.997)} & 0.900 (0.745-0.984) \\
        \rowcolor{gray!15}
        Myocarditis & 0.862 (0.698-0.984) & 0.917 (0.857-0.949) & 0.829 (0.683-0.944) & \textbf{0.937 (0.856-0.995)} \\
        \bottomrule
      \end{tabular}
  \end{adjustbox}
  \caption{The comparative results of CVDs diagnostic performance between the cardiac specialist diagnostic model and three SOTA models are presented in the form of AUC (95\% CI).}
  \label{AUC}
\end{table*}

\begin{table*}[htbp]
  \centering
  \renewcommand{\arraystretch}{1.2}
  \begin{adjustbox}{width=\textwidth,center}
      \begin{tabular}{@{}lcccc@{}}
        \toprule
        \rowcolor{gray!15}
        \multicolumn{5}{c}
        {\textbf{Sensitivity (Specificity=0.9)}} \\
        \multicolumn{5}{c}{\textbf{Internal Test Set}} \\
        \midrule
        & \textbf{Ours} & \textbf{VST} & \textbf{ViT} & \textbf{ResNet} \\
        \midrule
        \rowcolor{gray!15}
        NH & \textbf{0.938 (0.887-1.000)} & 0.923 (0.873-1.000) & 0.923 (0.876-1.000) & 0.953 (0.879-1.000) \\
        IHD & \textbf{0.839 (0.753-0.900)} & 0.776 (0.669-0.858) & 0.820 (0.710-0.890) & 0.739 (0.605-0.870) \\
        \rowcolor{gray!15}
        NICM & \textbf{0.850 (0.736-0.924)} & 0.750 (0.677-0.849) & 0.800 (0.704-0.889) & 0.793 (0.712-0.854) \\
        \midrule[0.5pt]
        HCM & 0.889 (0.822-0.985) & \textbf{0.931 (0.814-0.987)} & 0.877 (0.799-0.947) & 0.901 (0.813-1.000) \\
        \rowcolor{gray!15}
        DCM & \textbf{0.885 (0.793-1.000)} & 0.846 (0.682-1.000) & \textbf{0.885 (0.774-1.000)} & 0.808 (0.725-1.000) \\
        RCM & \textbf{0.889 (0.750-1.000)} & 0.778 (0.650-1.000) & 0.778 (0.631-1.000) & 0.873 (0.725-1.000) \\
        \rowcolor{gray!15}
        ACM & \textbf{0.769 (0.556-1.000)} & 0.764 (0.500-1.000) & 0.752 (0.462-1.000) & 0.758 (0.483-1.000) \\
        Myocarditis & \textbf{0.808 (0.652-0.957)} & 0.769 (0.594-0.939) & 0.778 (0.644-0.957) & 0.792 (0.630-0.964) \\
        \midrule[1.5pt]
        
        \multicolumn{5}{c}{\textbf{External Validation Set}} \\
        \midrule
        & \textbf{Ours} & \textbf{VST} & \textbf{ViT} & \textbf{ResNet} \\
        \midrule
        \rowcolor{gray!15}
        NH & 0.746 (0.571-0.898) & 0.791 (0.658-0.932) & 0.687 (0.541-0.815) & \textbf{0.881 (0.717-0.984)} \\
        IHD & \textbf{0.531 (0.340-0.744)} & 0.449 (0.256-0.610) & 0.325 (0.271-0.575) & 0.429 (0.214-0.564) \\
        \rowcolor{gray!15}
        NICM & \textbf{0.654 (0.412-0.719)} & 0.516 (0.389-0.637) & 0.428 (0.159-0.571) & 0.566 (0.446-0.675) \\
        \midrule[0.5pt]
        HCM & \textbf{0.721 (0.556-0.870)} & 0.587 (0.497-0.760) & 0.667 (0.468-0.835) & 0.714 (0.675-0.821) \\
        \rowcolor{gray!15}
        DCM & 0.714 (0.551-0.813) & \textbf{0.758 (0.649-0.849)} & 0.703 (0.527-0.828) & 0.717 (0.649-0.843) \\
        RCM & 0.333 (0.141-0.733) & 0.333 (0.153-0.616) & \textbf{0.500 (0.214-0.865)} & \textbf{0.500 (0.189-0.821)} \\
        \rowcolor{gray!15}
        ACM & 0.625 (0.250-1.000) & 0.625 (0.167-1.000) & \textbf{0.750 (0.599-1.000)} & 0.625 (0.285-1.000) \\
        Myocarditis & \textbf{0.727 (0.499-1.000)} & \textbf{0.727 (0.556-1.000)} & 0.455 (0.125-0.833) & 0.712 (0.462-1.000) \\
        \bottomrule
      \end{tabular}
  \end{adjustbox}
  \caption{The comparative results of CVDs diagnostic performance between the cardiac specialist diagnostic model and three SOTA models are presented in the form of Sensitivity (95\% CI).}
  \label{sen}
\end{table*}

\begin{table*}[htbp]
  \centering
  \renewcommand{\arraystretch}{1.2}
  \begin{adjustbox}{width=\textwidth,center}
      \begin{tabular}{@{}lcccc@{}}
        \toprule
        \rowcolor{gray!15}
        \multicolumn{5}{c}
        {\textbf{Specificity (Sensitivity=0.9)}} \\
        \multicolumn{5}{c}
        {\textbf{Internal Test Set}} \\
        \midrule
        & \textbf{Ours} & \textbf{VST} & \textbf{ViT} & \textbf{ResNet} \\
        \midrule
        \rowcolor{gray!15}
        NH & 0.954 (0.905-1.000) & 0.953 (0.900-1.000) & 0.950 (0.903-1.000) & \textbf{0.964 (0.901-1.000)} \\
        IHD & \textbf{0.838 (0.722-0.917)} & 0.824 (0.717-0.880) & 0.828 (0.711-0.900) & 0.819 (0.726-0.891) \\
        \rowcolor{gray!15}
        NICM & \textbf{0.862 (0.808-0.933)} & 0.818 (0.740-0.878) & 0.769 (0.696-0.897) & 0.756 (0.692-0.841) \\
        HCM & 0.904 (0.836-0.989) & \textbf{0.940 (0.859-1.000)} & 0.902 (0.860-0.989) & 0.900 (0.831-1.000) \\
        \rowcolor{gray!15}
        DCM & \textbf{0.946 (0.821-0.985)} & 0.891 (0.830-0.985) & 0.938 (0.855-0.993) & 0.930 (0.882-0.984) \\
        RCM & 0.942 (0.870-0.985) & 0.934 (0.813-0.980) & 0.861 (0.745-0.986) & \textbf{0.963 (0.887-0.993)} \\
        \rowcolor{gray!15}
        ACM & 0.845 (0.775-1.000) & 0.845 (0.775-1.000) & \textbf{0.880 (0.739-1.000)} & \textbf{0.880 (0.548-1.000)} \\
        Myocarditis & \textbf{0.853 (0.603-1.000)} & 0.713 (0.618-1.000) & 0.752 (0.569-1.000) & 0.767 (0.580-1.000) \\
        \midrule[1.5pt]
        
        \multicolumn{5}{c}{\textbf{External Validation Set}} \\
        \midrule
        & \textbf{Ours} & \textbf{VST} & \textbf{ViT} & \textbf{ResNet} \\
        \midrule
        \rowcolor{gray!15}
        NH & 0.861 (0.771-0.912) & 0.865 (0.817-0.924) & 0.740 (0.665-0.809) & \textbf{0.899 (0.854-0.946)} \\
        IHD & \textbf{0.589 (0.339-0.768)} & 0.513 (0.277-0.758) & 0.392 (0.246-0.598) & 0.580 (0.371-0.745) \\
        \rowcolor{gray!15}
        NICM & 0.405 (0.307-0.514) & 0.517 (0.400-0.643) & 0.483 (0.357-0.600) & \textbf{0.552 (0.438-0.689)} \\
        HCM & \textbf{0.698 (0.607-0.875)} & 0.653 (0.598-0.755) & 0.685 (0.589-0.841) & 0.628 (0.523-0.766) \\
        \rowcolor{gray!15}
        DCM & 0.600 (0.242-0.816) & 0.632 (0.323-0.806) & \textbf{0.647 (0.486-0.821)} & 0.544 (0.386-0.781) \\
        RCM & \textbf{0.863 (0.697-0.930)} & 0.771 (0.548-0.961) & 0.588 (0.483-0.968) & 0.830 (0.643-0.961) \\
        \rowcolor{gray!15}
        ACM & 0.815 (0.575-0.981) & 0.874 (0.447-0.954) & \textbf{0.954 (0.442-1.000)} & 0.815 (0.642-0.948) \\
        Myocarditis & 0.607 (0.397-0.986) & 0.704 (0.540-1.000) & 0.547 (0.299-0.913) & \textbf{0.851 (0.583-1.000)} \\
        \bottomrule
      \end{tabular}
  \end{adjustbox}
  \caption{The comparative results of CVDs diagnostic performance between the cardiac specialist diagnostic model and three SOTA models are presented in the form of Specificity (95\% CI).}
  \label{spc}
\end{table*}

\begin{figure*}[!t]
	\centering
		\includegraphics[width=1 \textwidth] {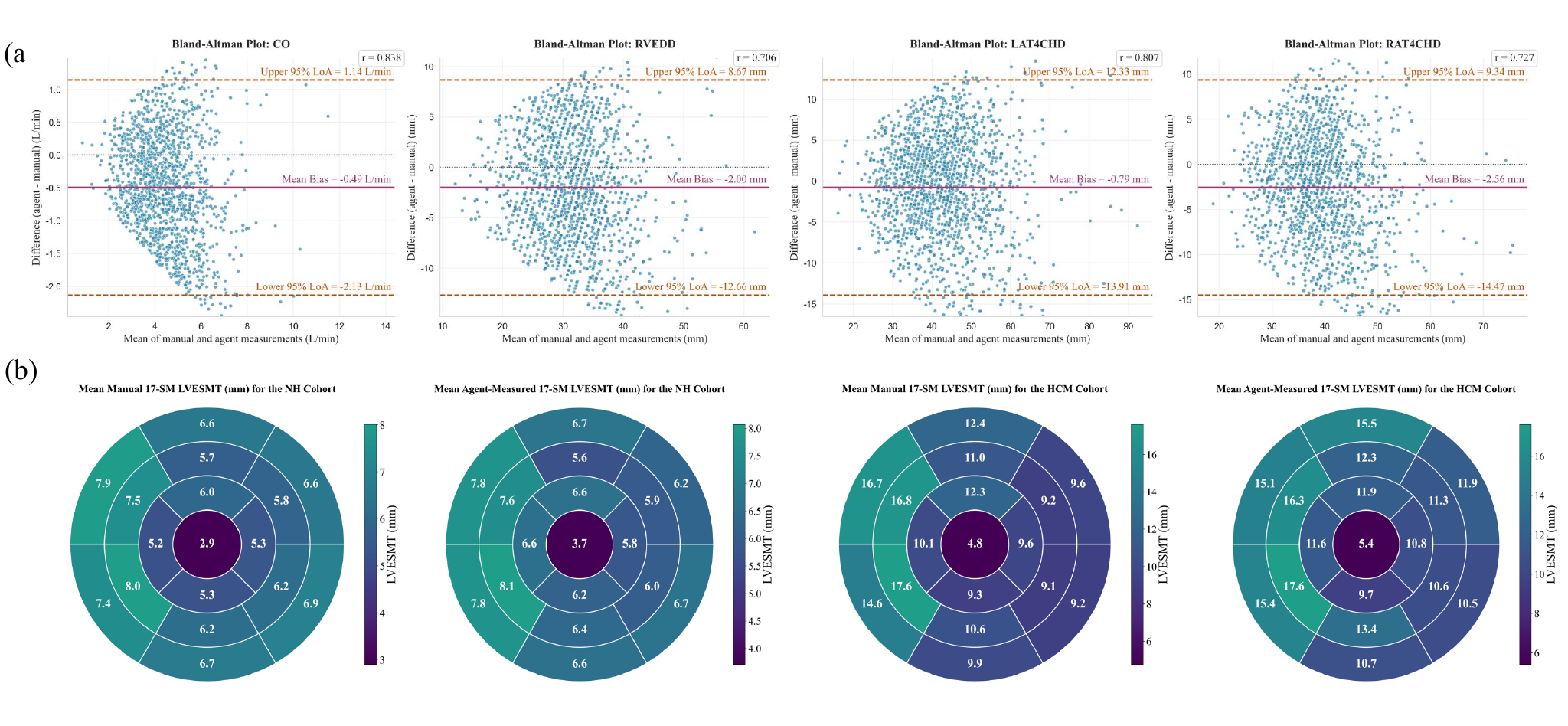} 
	\caption{Consistency evaluation between the BAAI Cardiac Agent and clinical reports. (a) Bland-Altman plots compare cardiac output (CO), right ventricular end-diastolic diameter (RVEDD), and left/right atrial maximum transverse diameters (LAT4CHD, RAT4CHD) measured by the BAAI Cardiac Agent with manual clinical reports in the internal cohort; (b) Mean left ventricular end-diastolic wall thickness (LVEDWT) distributions from manual 17-segment model (17-SM) reports and the BAAI Cardiac Agent in internal normal and hypertrophic cardiomyopathy cohorts. The bullseye plot sequentially maps left ventricular basal, mid, and apical segments to the plot's outer, middle, and inner layers, with the center representing the apex.
    } \label{s2}
\end{figure*}

\begin{figure*}[!t]
	\centering
		\includegraphics[width=1 \textwidth] {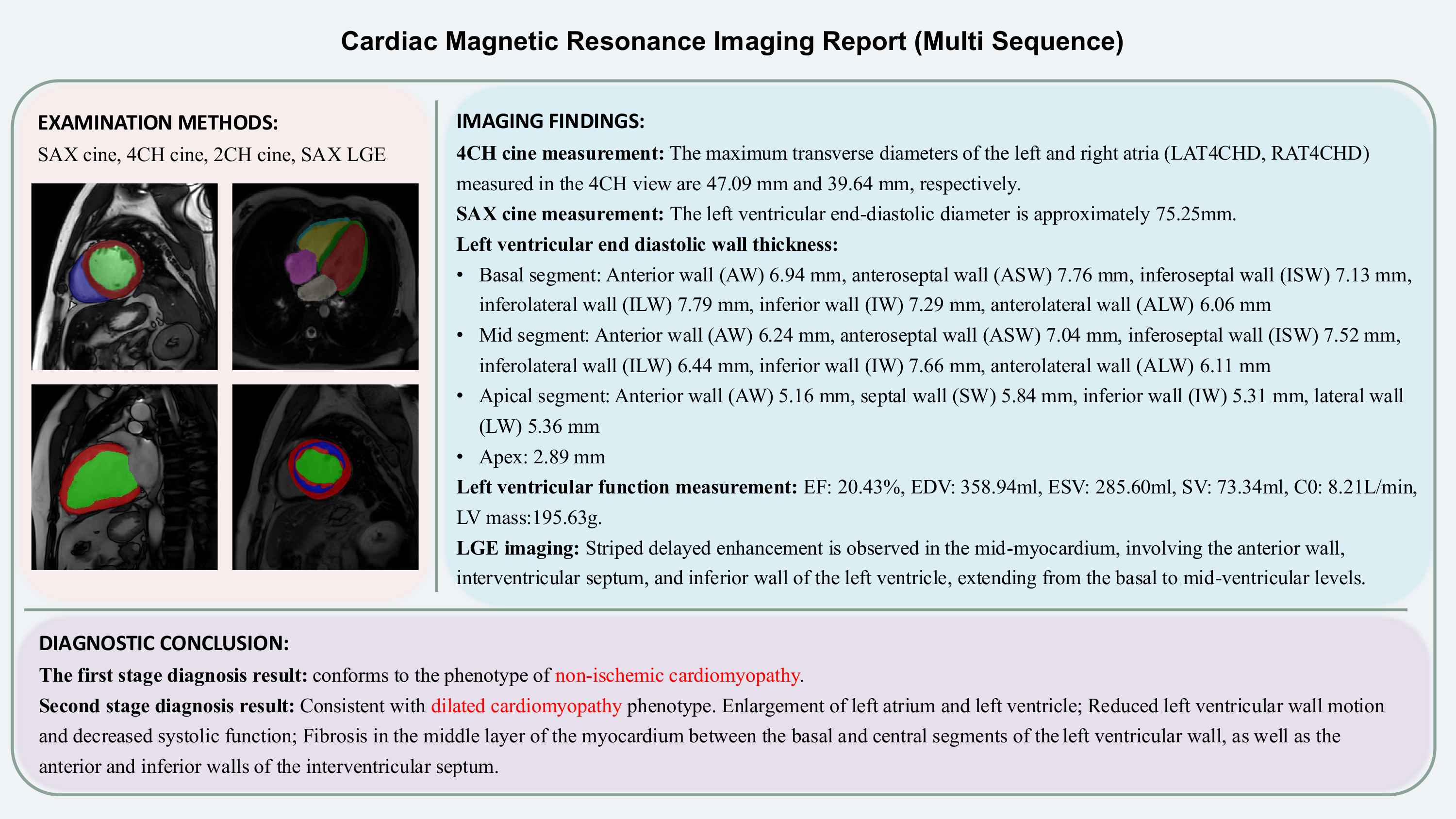} 
	\caption{Schematic diagram of a complete medical report generated by the BAAI Cardiac Agent, illustrating the measurement of cardiac function parameters and imaging diagnostic analysis for a patient with dilated cardiomyopathy (DCM).
    } \label{s3}
\end{figure*}

\begin{landscape}
\begin{figure*}[!t]
	\centering
	\includegraphics[width=1.5 \textwidth] {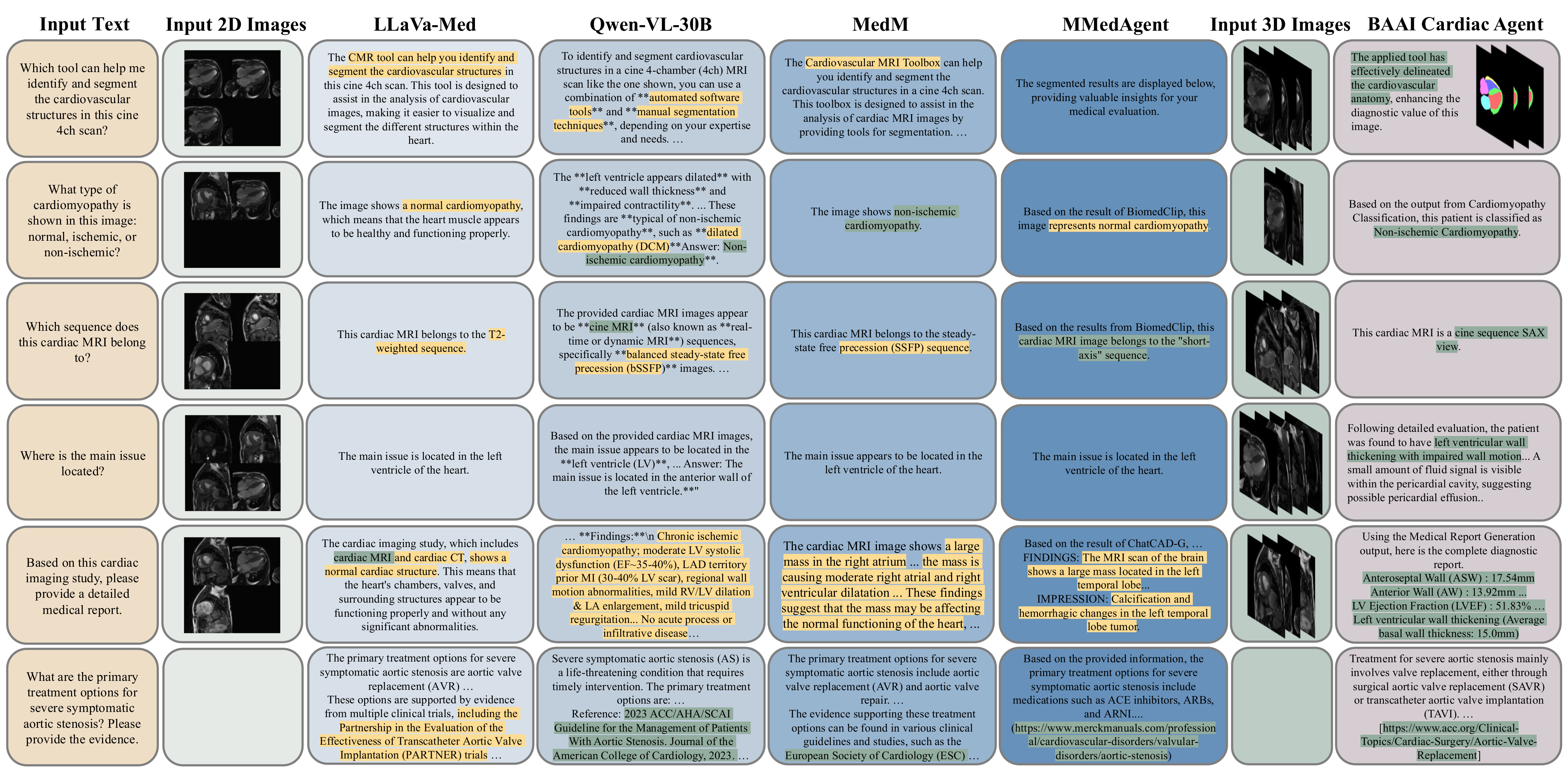} 
	\caption{Qualitative comparison between BAAI Cardiac Agent and other SOTA methods across different tasks. Undesired and desired responses are highlighted in licorice yellow and pine green, respectively.
    } \label{s4}
\end{figure*}
\end{landscape}



\end{appendices}
\end{document}